# Vibrations and waves in soft dielectric elastomer structures[#]

Zinan Zhao[1], Yingjie Chen[1], Xueyan Hu[1], Ronghao Bao[1], Bin Wu[2*], Weiqiu Chen[1,3,4*]

[1]Key Lab of CAD & CG, Key Laboratory of Soft Machines and Smart Devices of Zhejiang Province & Department of Engineering Mechanics, Zhejiang University, Hangzhou 310027, P. R. China

[2]School of Mathematical and Statistical Sciences, University of Galway, University Road, Galway, Ireland

[3]Soft Matter Research Center, Zhejiang University, Hangzhou 310027, P. R. China

[4]Shenzhen Research Institute of Zhejiang University, Shenzhen 518057, P.R. China

**Abstract**

Dielectric elastomers (DEs) are a type of multifunctional materials with salient features that are very attractive in developing soft, lightweight, and small-scale transducers and robotics. This paper reviews the mechanics of soft DE structures, focusing on their dynamic responses, including vibrations and waves in typical structures. The basic formulations of nonlinear electroelasticity are briefly summarized by following the descriptions of Dorfmann and Ogden. The recent advances in the study on linearized dynamic problems of DE plates and shells are then outlined according to the geometric configurations as well as the coordinates thereof adopted. Some interesting observations from the numerical results are highlighted, and the underlying physics is discussed. For completeness, an account of the nonlinear dynamic characteristics of DE structures is also given. For DE structures with a periodic arrangement, i.e., DE phononic crystals (PCs) or metamaterials, this paper reviews the designs, solutions, and results to illustrate the wide-range tunability of band structure that is realized by large reversible deformation and electromechanical coupling. The present literature survey could be a useful guide for future studies on the dynamic topics of DE structures.

---

[#] This paper is dedicated to Professor Marian Wiercigroch on the occasion of his sixtieth birthday.
[*] Corresponding authors. Tel./Fax: 86-571-87951866; E-mail: chenwq@zju.edu.cn (WQC); wubinlongchang@163.com (BW).

**Catalog**

# 1. Introduction

Materials featuring multifield couplings have been widely used in developing functional structures and devices due to their capability of transforming energy from one kind to another (e.g., from mechanical to electrical). Among these functional materials, piezoelectric materials exhibiting the electromechanical coupling are the most popular one that could be easily found even in everyone's daily life. However, the conventional piezoelectric materials, such as quartz and ceramics, are very brittle, and hence they are unable to withstand large deformation. New applications, such as soft robotics and flexible electronics, call for materials that can deform significantly. Dielectric elastomers (DEs) are such a type of multifunctional materials that are very ideal for developing soft, lightweight, and small-scale transducers, actuators, sensors, energy harvesters, and soft robotics [1-8]. For example, the DE generators have been proven both numerically and experimentally to be a very promising energy conversion technology, workable in different dimensional ranges [1]. Moreover, the applied electric stimuli significantly modify the effective electromechanical properties and geometric configurations of the DE materials/structures, resulting in changes of their mechanical characteristics. Consequently, the DE structures also have various potential applications including electrostatically tunable resonators, oscillators, waveguides, phononic crystals (PCs), metamaterials, and some components for acoustics and vibration control [9-13].

Designing practically reliable DE devices and structures relies on the understanding of their mechanical behavior under both static and dynamic stimuli, which requires a well-established nonlinear continuum theory. The two prominent features of DEs, i.e., electromechanical coupling and large deformation, have already been properly described in the nonlinear continuum theory of elastic dielectrics that was proposed by Toupin [14,15] almost sixty years ago. With the development of novel soft DE devices and structures, this nonlinear theoretical framework has been reformulated and also refined in different forms in recent decades. The reader is referred to a recent review article for the comprehensive comparison between different versions of the theory of nonlinear electroelasticity by Wu et al. [16].

The advances in the study of DE materials and structures have been well teased and summarized from different aspects but all from the mechanics point of view. Suo [17] gave an elegant but incomparably clear account of the mechanical theory of DEs, with emphases on its connections with empirical observations, molecular pictures, and applications. Liu et al. [18] paid their attention to both DEs and their composites as well as some typical applications, including actuators, artificial muscles, insect robots, energy harvesters, and Braille tactile displays. The survey by Zhu et al. [19] focuses on the theoretical and numerical studies that outline the existing approaches and methods in the mechanical analysis of DEs, aiming at sorting out the important issues for improving the performance and reliability of DE structures and devices. Dorfmann and Ogden [20] presented a rather complete exposition of the study on DEs and DE structures with particular reference to the theoretical framework of nonlinear electroelasticity proposed by themselves [21-23]. Lu et al. [24] reported a summary of the recent advances in the theory of DEs and demonstrated some examples of using the theory to design DE transducers. In contrast to the above-mentioned review papers, which all are inclined toward the constitutive behavior of DEs and the static responses of DE structures, Huang et al. [25] summarized the state-of-the-art of dynamic responses and vibration controls of DE structures. The analytical models presented in Huang et al. [25] are all approximate, not formulated directly in the three-dimensional (3D) setting.

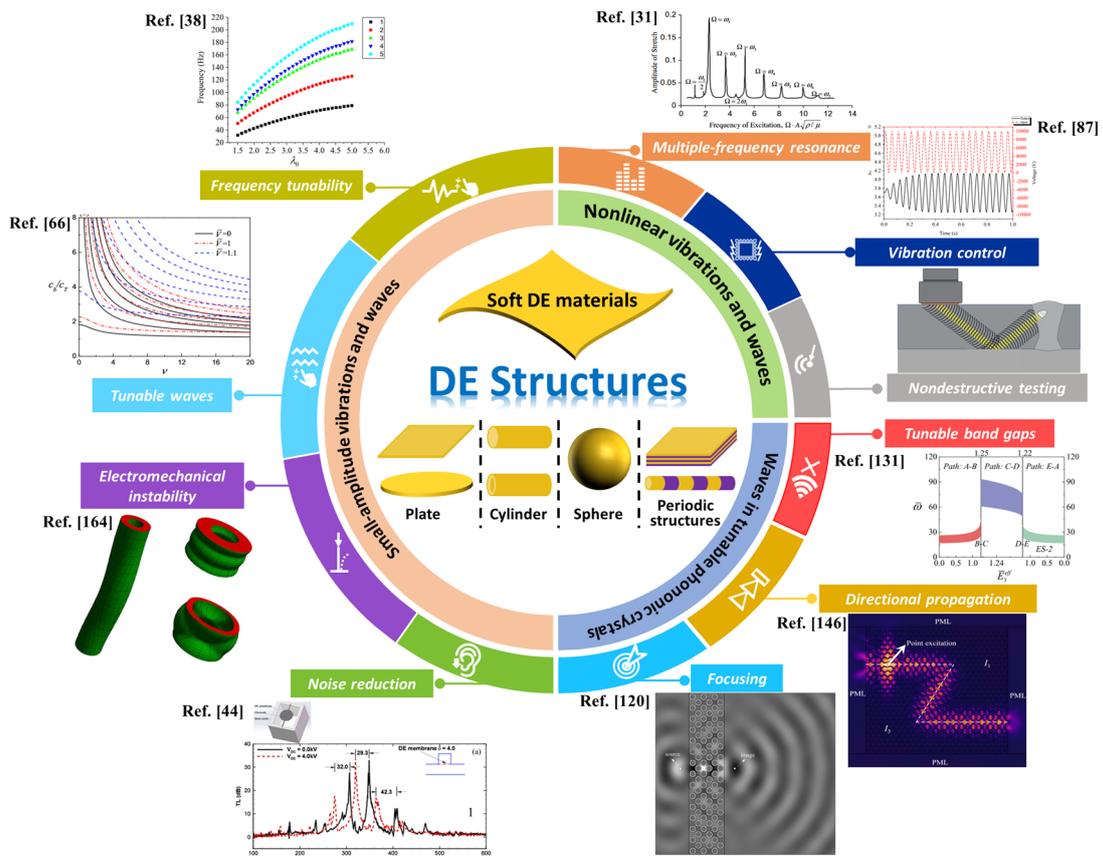

**Figure 1**. The outline of this review. (Reproduced with permission from Zhu et al. [31], copyright 2010 by Elsevier; Reproduced with permission from Li et al. [38], copyright 2018 by IOP Publishing; Reproduced with permission from Lu et al. [44], copyright 2015 Acoustical Society of America; Reproduced with permission from Wu et al. [66], copyright 2016 Elsevier; Reproduced with permission from Li et al. [87], copyright 2018 Elsevier; Reproduced with permission from Yang et al. [120], copyright 2008 IOP Publishing; Reproduced with permission from Chen et al. [131], copyright 2020 Elsevier; Reproduced with permission from Zhou et al. [146], copyright 2019 Elsevier; Reproduced with permission from Su et al. [164], copyright 2016 Elsevier)

The present survey, while again from the mechanics viewpoint, intends to show the advances in the study of vibrations and waves in DE structures. Following a brief introduction to the theory of nonlinear electroelasticity formulated by Dorfmann and Ogden [21-23], we pay our attention first to the 3D linearized dynamic analyses of DE plates and shells. An efficient method based on the state-space formulations in different coordinates is particularly mentioned. Numerical results are selectively shown and the underlying physical mechanisms are discussed. Then, we also provide an updated account of the nonlinear dynamic analyses of DE structures, which could serve as a supplement to the earlier review of Huang et al. [25]. Our last focus point here is DE phononic crystals or metamaterials, which exhibit a certain kind of periodicity in structural geometry, boundary conditions, or/and material properties. We pay particular attention to some recently reported designs to show the analytical solutions (if available) and the results to demonstrate that large deformation and electromechanical coupling, which are the two intrinsic properties of DEs, could be utilized to manipulate wave propagation in DE PCs/metamaterials. This survey, including more than 180 references, most of which are published in recent years, is believed to be a timely and informative guide for future studies on DE structures and devices with particular reference to their dynamic responses. The outline of this survey is illustrated in **Fig. 1**.

## 2. Nonlinear electroelasticity

As mentioned earlier, there are seemingly quite different versions of the nonlinear continuum theory for a deformable body with electromechanical coupling. A detailed comparison by Wu et al. [16] indicates that all these versions, if correctly derived, are equivalent to each other. Here we follow the descriptions of Dorfmann and Ogden [21-23,26] to give a brief introduction to the theory.

We use the three configurations (i.e., reference $B_r$, initial $\tilde{B}$, and current $B$) in **Fig. 2** to identify the three stages of the motion of a deformable body [16]. At the first stage the DE body, free from stress and electric displacement, is in the undeformed reference configuration; at the second stage, the body deforms statically under certain stimuli and takes the initial configuration; at the third stage, the body further deforms with respect to the initial configuration and now takes the current configuration. For static deformation without instability, we could just consider the first two stages, while for buckling, vibration, or wave propagation analysis, we usually need to consider all three stages so that the effect of biasing fields (i.e., the pre-deformation and other accompanying electromechanical fields) could be examined. In the following, the general nonlinear theory that applies to the initial static finite deformation (from the first stage to the second stage, characterized by the initial displacement $\tilde{\mathbf{u}}$ and other accompanying physical fields) and the linearized theory for the incremental dynamic deformation (from the second stage to the third stage, with the incremental displacement $\mathbf{u}$ and other accompanying physical fields as well, which are assumed to be infinitesimal), are presented in sequence.

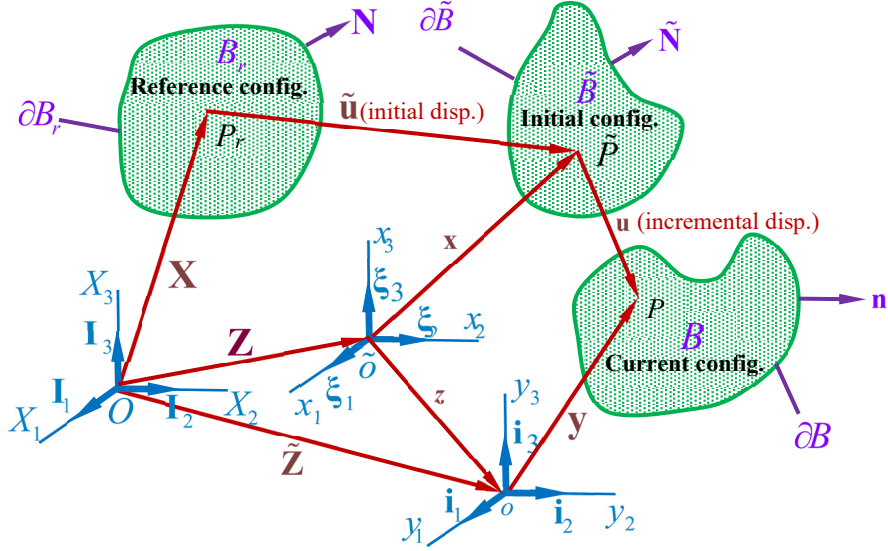

**Figure 2**. Reference configuration, initial configuration, and current configuration of a deformable body. Three Cartesian coordinate systems $(X_1, X_2, X_3)$, $(x_1, x_2, x_3)$, and $(y_1, y_2, y_3)$ are used to describe the three configurations, with $(\mathbf{I}_1, \mathbf{I}_2, \mathbf{I}_3)$, $(\boldsymbol{\xi}_1, \boldsymbol{\xi}_2, \boldsymbol{\xi}_3)$, $(\mathbf{i}_1, \mathbf{i}_2, \mathbf{i}_3)$ being the corresponding orthogonal sets of the unit base vectors, respectively. $\partial B_r$, $\partial \tilde{B}$, and $\partial B$ denote the boundaries, while $\mathbf{N}$, $\tilde{\mathbf{N}}$, and $\mathbf{n}$ denote the outward unit normal vectors of the body in the three configurations, respectively. $\mathbf{Z}$, $\tilde{\mathbf{Z}}$, and $\mathbf{z}$ are the relative position vectors between the origins of the coordinates as evidenced in the figure. $\tilde{\mathbf{u}}$ and $\mathbf{u}$ are the initial and incremental displacement vectors, respectively.

## 2.1 General theory of nonlinear electroelasticity

As shown in **Fig. 2**, we denote the boundary of the body in the undeformed reference configuration as $\partial B_r$, with the outward normal being $\mathbf{N}$. A material point at $P_r$, which is identified with its position vector $\mathbf{X}$ in the reference configuration, moves to a new point $\tilde{P}$ in the initial configuration with the position vector $\mathbf{x}$ after a static finite deformation described by the smooth tensorial transformation $\mathbf{x} = \boldsymbol{\chi}(\mathbf{X})$. The boundary of the deformed body in the initial configuration is denoted as $\partial \tilde{B}$ with the outward normal $\tilde{\mathbf{N}}$. The equations of equilibrium in the generalized sense can be written as

$$\text{Div}\,\mathbf{T} = \mathbf{0}, \quad \text{Curl}\,\mathbf{E}_l = \mathbf{0}, \quad \text{Div}\,\mathbf{D}_l = 0, \tag{1}$$

where the divergence operator (Div) and the curl operator (Curl) [as well as the gradient operator (Grad) to appear immediately] are all with respect to the undeformed reference configuration $B_r$, $\mathbf{T} = J\mathbf{F}^{-1}\boldsymbol{\tau}$ is the total nominal stress tensor ($J = \det \mathbf{F}$), and $\mathbf{E}_l = \mathbf{F}^T\mathbf{E}$ and $\mathbf{D}_l = J\mathbf{F}^{-1}\mathbf{D}$ are the Lagrangian counterparts (and hence with subscript $l$) of the true electric field vector $\mathbf{E}$ and electric displacement vector $\mathbf{D}$, respectively. Here $\boldsymbol{\tau}$ denotes the total Cauchy stress tensor (with respect to the initial configuration), and $\mathbf{F} = \text{Grad}\,\chi$ is the deformation gradient tensor. The 'quasi-electrostatic approximation' for the electric field has been assumed so that no magnetic field is involved in Eq. (1). By assuming the existence of an energy density function $\Omega = \Omega(\mathbf{F}, \mathbf{D}_l)$ per unit volume in $B_r$, we may express the constitutive relations for a general compressible DE as

$$\mathbf{T} = \frac{\partial \Omega}{\partial \mathbf{F}}, \quad \mathbf{E}_l = \frac{\partial \Omega}{\partial \mathbf{D}_l}. \tag{2}$$

Since the body is assumed to be free from stress and electric displacement in $B_r$, we have $\Omega(\mathbf{I}, \mathbf{0}) = 0$. For an isotropic DE, we can obtain

$$J\boldsymbol{\tau} = 2\Omega_1 \mathbf{b} + 2\Omega_2(I_1\mathbf{b} - \mathbf{b}^2) + 2I_3\Omega_3\mathbf{I} + 2\Omega_5 I_3 \mathbf{D} \otimes \mathbf{D} + 2\Omega_6 I_3 (\mathbf{D} \otimes \mathbf{bD} + \mathbf{bD} \otimes \mathbf{D}),$$

$$\mathbf{E} = 2J(\Omega_4 \mathbf{b}^{-1}\mathbf{D} + \Omega_5 \mathbf{D} + \Omega_6 \mathbf{bD}), \tag{3}$$

where $\Omega_m = \partial \Omega / \partial I_m$ ($m = 1, 2, \cdots, 6$), and $I_1 = \text{tr}\,\mathbf{c}$, $I_2 = \frac{1}{2}[(\text{tr}\,\mathbf{c})^2 - \text{tr}(\mathbf{c}^2)]$, $I_3 = \det \mathbf{c}$, $I_4 = \mathbf{D}_l \cdot \mathbf{D}_l$, $I_5 = \mathbf{D}_l \cdot (\mathbf{c}\mathbf{D}_l)$ and $I_6 = \mathbf{D}_l \cdot (\mathbf{c}^2 \mathbf{D}_l)$ are the six scalar invariants. Here, $\mathbf{b} = \mathbf{FF}^T$ and $\mathbf{c} = \mathbf{F}^T\mathbf{F}$ are the left and right Cauchy-Green tensors, respectively.

The boundary conditions may be described as

$$\boldsymbol{\tau}\tilde{\mathbf{N}} = \mathbf{t}_a + \mathbf{t}_e, \quad (\mathbf{E} - \mathbf{E}^*) \times \tilde{\mathbf{N}} = \mathbf{0}, \quad (\mathbf{D} - \mathbf{D}^*) \cdot \tilde{\mathbf{N}} = 0 \tag{4}$$

where $\mathbf{t}_a$ is the applied mechanical traction per unit area of $\partial \tilde{B}$, and $\mathbf{t}_e$ is to be determined by considering the effect of the exterior electric field ($\mathbf{E}^*$, $\mathbf{D}^*$) in the medium surrounding the DE body. In vacuum, the electric field is governed by

$$\text{div}\,\mathbf{D}^* = 0, \quad \text{curl}\,\mathbf{E}^* = \mathbf{0} \tag{5}$$

where $\mathbf{D}^* = \varepsilon_0 \mathbf{E}^*$, with $\varepsilon_0$ being the permittivity of free space, and the divergence operator (div) and the curl operator (curl) [as well as the gradient operator (grad) to appear in the following] are all with respect to the initial configuration $\tilde{B}$.

If the material is incompressible (i.e., $J=1$), Eqs. (2) and (3) will be replaced by

$$\mathbf{T} = \frac{\partial \Omega}{\partial \mathbf{F}} - p\mathbf{F}^{-1}, \quad \mathbf{E}_l = \frac{\partial \Omega}{\partial \mathbf{D}_l} \tag{6}$$

and

$$\boldsymbol{\tau} = 2\Omega_1 \mathbf{b} + 2\Omega_2 (I_1 \mathbf{b} - \mathbf{b}^2) - p\mathbf{I} + 2\Omega_5 \mathbf{D} \otimes \mathbf{D} + 2\Omega_6 (\mathbf{D} \otimes \mathbf{bD} + \mathbf{bD} \otimes \mathbf{D})$$

$$\mathbf{E} = 2(\Omega_4 \mathbf{b}^{-1}\mathbf{D} + \Omega_5 \mathbf{D} + \Omega_6 \mathbf{bD}) \tag{7}$$

respectively, where $p$ is the Lagrange multiplier associated with the constraint of material incompressibility.

### 2.2 Linearized theory of infinitesimal incremental fields

With respect to the initial configuration, the DE body further moves with a time-dependent incremental displacement $\mathbf{u}$, which is assumed to be infinitesimal. Then, the governing equations and boundary conditions for the incremental fields (a superimposed dot will be used to signify the corresponding physical variables) could be derived from the above general nonlinear theory via linear perturbation. The results for a compressible DE are:

$$\operatorname{div} \dot{\mathbf{T}}_0 = \rho \frac{\partial^2 \mathbf{u}}{\partial t^2}, \quad \operatorname{curl} \dot{\mathbf{E}}_{l0} = \mathbf{0}, \quad \operatorname{div} \dot{\mathbf{D}}_{l0} = 0 \tag{8}$$

$$\dot{\mathbf{T}}_0 = \mathcal{A}_0 \mathbf{H} + \boldsymbol{\Gamma}_0 \dot{\mathbf{D}}_{l0}, \quad \dot{\mathbf{E}}_{l0} = \boldsymbol{\Gamma}_0^{\mathrm{T}} \mathbf{H} + \boldsymbol{K}_0 \dot{\mathbf{D}}_{l0}, \tag{9}$$

$$\dot{\mathbf{T}}_0^{\mathrm{T}} \mathbf{n} = \dot{\mathbf{t}}_{A0} + \dot{\boldsymbol{\tau}}^* \mathbf{n} - \boldsymbol{\tau}^* \mathbf{H}^{\mathrm{T}} \mathbf{n} + (\operatorname{div}\mathbf{u}) \boldsymbol{\tau}^* \mathbf{n},$$

$$(\dot{\mathbf{E}}_{l0} - \dot{\mathbf{E}}^* - \mathbf{H}^{\mathrm{T}} \mathbf{E}^*) \times \mathbf{n} = \mathbf{0}, \quad \left[ \dot{\mathbf{D}}_{l0} + \mathbf{H}\mathbf{D}^* - \dot{\mathbf{D}}^* - (\operatorname{div}\mathbf{u}) \mathbf{D}^* \right] \cdot \mathbf{n} = 0 \quad (\text{on } \partial B), \tag{10}$$

where $\rho = \rho_r / J$ is the current mass density with $\rho_r$ being that in the undeformed reference configuration, $\mathbf{H} = \operatorname{grad} \mathbf{u}$, and $\dot{\mathbf{T}}_0 = J^{-1} \mathbf{F} \dot{\mathbf{T}}$, $\dot{\mathbf{E}}_{l0} = \mathbf{F}^{-\mathrm{T}} \dot{\mathbf{E}}_l$, and $\dot{\mathbf{D}}_{l0} = J^{-1} \mathbf{F} \dot{\mathbf{D}}_l$ are the 'push forward' versions of $\dot{\mathbf{T}}$, $\dot{\mathbf{E}}_l$, and $\dot{\mathbf{D}}_l$, respectively. In Eq. (9), $\mathcal{A}_0$, $\boldsymbol{\Gamma}_0$ and $\boldsymbol{K}_0$ are the effective electroelastic moduli tensors, which depend on the biasing fields ($\mathbf{F}, \mathbf{D}_l$) [26]. In Eq. (10)$_1$, $\dot{\mathbf{t}}_{A0}$ is the incremental surface mechanical traction vector per unit area of $\partial B$.

For an incompressible DE, besides the constraint $\text{div}\,\mathbf{u} = 0$, the constitutive relations (9) should be replaced by

$$\dot{\mathbf{T}}_0 = \boldsymbol{\mathcal{A}}_0 \mathbf{H} + \boldsymbol{\varGamma}_0 \dot{\mathbf{D}}_{l0} + p\mathbf{H} - \dot{p}\mathbf{I}, \quad \dot{\mathbf{E}}_{l0} = \boldsymbol{\varGamma}_0^{\mathrm{T}} \mathbf{H} + \boldsymbol{K}_0 \dot{\mathbf{D}}_{l0}, \tag{11}$$

It is noted that, since the incremental motion is infinitesimal, the difference between the initial configuration and the current configuration could be neglected such that $\partial B$ and $\mathbf{n}$ can be regarded as identical with $\partial \tilde{B}$ and $\tilde{\mathbf{N}}$, respectively.

## 3. Small-amplitude vibrations and waves in deformed DE structures

As mentioned above, when the DE structure is subjected to a finite deformation, in addition to the geometric change, the effective electroelastic moduli also change with that deformation. Thus, the finite deformation of the DE structures induced by, say, the applied electric fields could be used to tune their dynamic responses of interest. The researches on superimposed small-amplitude vibrations and waves in soft elastic materials begin in the late 1950s, but most earlier works focus on waves [27-29]. In this section, we will summarize the recent advances in small-amplitude vibrations and waves in finitely deformed DE structures of a plate, cylindrical, or spherical configuration, and a summary of the relevant works is given in **Table 1**.

### 3.1 Rectangular and circular plates

The tunability of vibration responses (e.g., resonance frequencies and mode shapes) of micro-electro-mechanical system (MEMS) devices was first realized by Dubois et al. [30] by designing a circular DE polymer membrane made of polydimethylsiloxane (PDMS) and with ion-implanted electrodes. It was experimentally found that the resonance frequency of the DE membrane could be repeatedly and reversibly tuned up to 77% by applying a certain actuation voltage. This pioneering work opens the door to frequency-tunable MEMS devices made of soft DE materials. In this subsection, the relevant researches on small-amplitude vibrations and waves superimposed on finitely deformed soft DE plates of rectangular and circular configurations are reviewed.

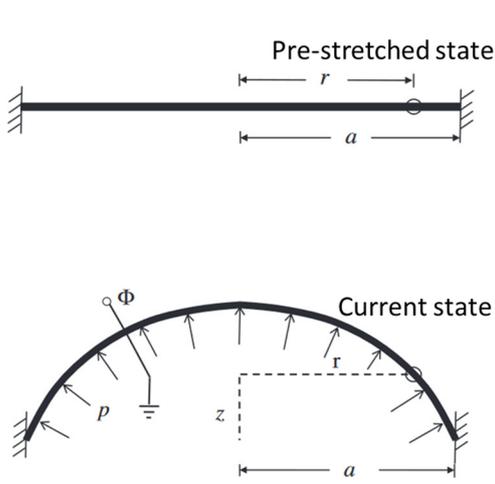
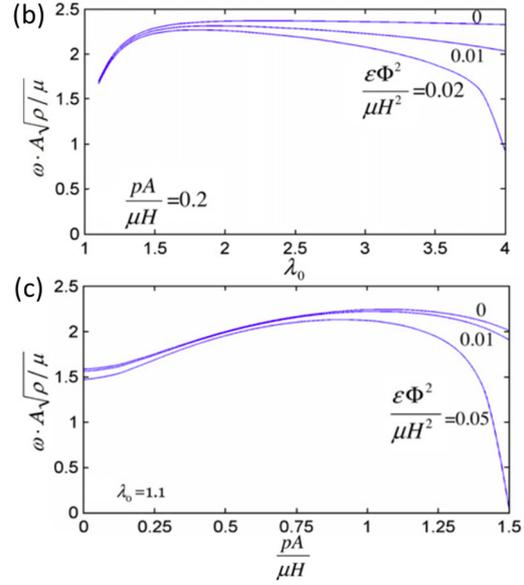

(a) Finitely deformed circular DE membrane

**Figure 3.** Small-amplitude vibrations of a finitely deformed circular DE membrane[31]: (a) A finitely deformed DE membrane under in-plane pre-stretch and out-of-plane pressure; (b) variation of the fundamental frequency $\omega \cdot A\sqrt{\rho/\mu}$ with the pre-stretch $\lambda_0$ for a fixed pressure $pA/\mu H$ and several applied voltages $\varepsilon\Phi^2/\mu H^2$; (c) variation of the fundamental frequency with the pressure for a fixed pre-stretch and several applied voltages. (Reproduced with permission from Zhu et al. [31], copyright 2010 by Elsevier)

    Different from the in-plane finite deformations in Ref. [30], Zhu et al. [31] analytically and experimentally investigated the small-amplitude vibrations superimposed on a circular ideal DE membrane considered by Zhao et al. [32] with both in-plane and out-of-plane finite deformations, which were achieved by the in-plane initial stretch and the out-of-plane pressure, respectively (see **Fig. 3**(a)). They found that the fundamental resonance frequency of the circular DE membrane is tunable by modulating the in-plane pre-stretch, out-of-plane pressure, and applied voltage (see **Fig. 3**(b) and (c)). Moretti et al. [33] also designed an out-of-plane pre-stretched circular annular DE membrane, described as an incompressible visco-hyperelastic material [34], via the out-of-plane forces applied by a spring connected to a rigid solid disc at the center of the membrane, and then they investigated free and forced vibrations of the DE structure subjected to the external voltages numerically and experimentally. The experimental results illustrated that the circular annular membrane vibrates with a pistonic mode at low excitation frequencies, while complex modes, e.g., circumferential and axial-symmetrical modes, are observed at higher excitation frequencies. The appearance of these vibration modes can be controlled by changing the out-of-plane

initial deformation. Moreover, they proposed a fully-coupled finite-element model, which was validated by the experimental results, to help design multi-functional actuators with selective excitation of target vibration modes. Chakravarty [35] paid distinctive attention to the surrounding air effect on vibration responses of the deformed circular DE plates, which are characterized by a Mooney-Rivlin hyperelastic material, by using the developed analytical and finite element models. The simulation results illustrated that the damping effect of air, although reducing the out-of-plane vibration amplitudes, has a neglectable effect on the resonance frequencies, while the increase in the voltage and electrode mass significantly decreases the frequencies. Also, Feng et al. [36] developed an analytical model to investigate the influences of ambient pressure and applied voltage on the vibration properties of a double-clamped DE microbeam resonator based on the Euler-Bernoulli beam model and the squeeze-film theory. The numerical results showed that increasing the applied voltage enhances the frequency shift ratio and then improves the sensitivity of the resonator. However, when the voltage approaches the cut-off value of buckling, the frequency shift increases dramatically, which indicates that the resonator becomes too unstable to control the frequency tuning. Hence, the external voltage should be applied in a reasonable range to avoid instability and failure of the resonator. Subsequently, Li et al. [37] analytically investigated the dynamic responses of the in-plane pre-stretched rectangular Gent DE resonator, which consists of a passive DE membrane and an active DE membrane, with electric loads applied on the active one (see **Fig. 4**(a)). The obtained results in **Fig. 4**(b) and (c) revealed that the constant voltage and pre-stretch provide an effective means to manipulate the natural frequencies of the resonator, while they also emphasized the safe range of applied voltages to avoid mechanical instability and failure of the resonator. Furthermore, Li et al. [38] designed a two-layered system formed by attaching a stiffer passive membrane (neo-Hookean (NH) material) onto the active circular DE membrane (ideal DE material in [32]), and an external voltage was applied to the active one to tune the vibration properties (see **Fig. 4**(d)). They found that the resonance frequencies are not only dependent on the material properties and geometric size of the passive layer (see **Fig. 4**(f)) but also can be actively tuned by the applied voltage and pre-stretch (see **Fig. 4**(e)). In addition, based on the springs with variable stiffness, Dong et al. [39] developed a lumped model to predict the vibration responses of a pre-stretched circular

Yeoh membrane under external voltage and validated the analytical model through the experimental and finite element results.

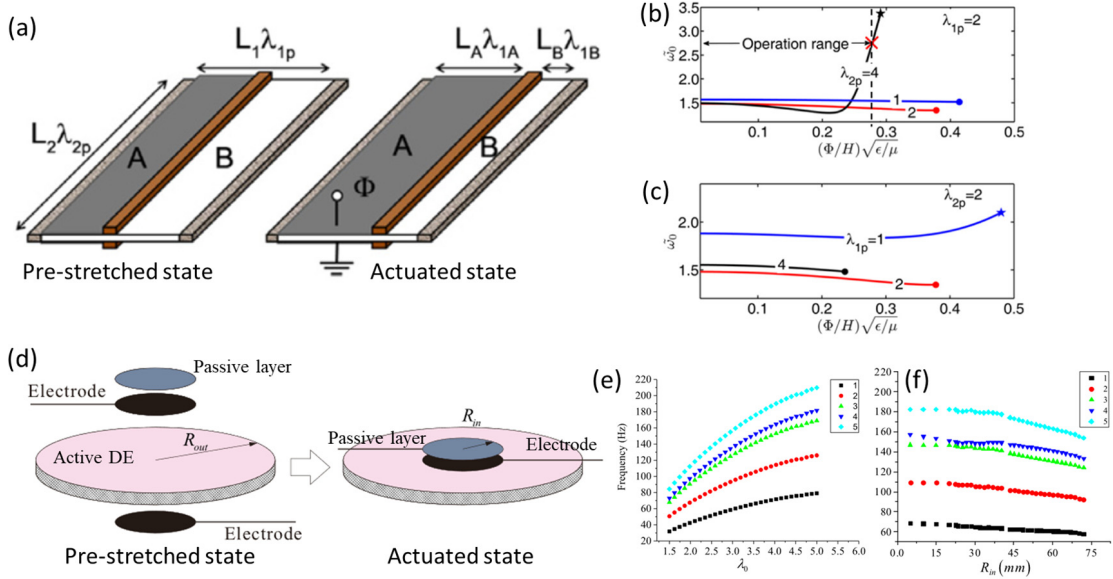

**Figure 4.** Vibration responses of finitely deformed DE structures with passive and active membranes[37,38]: (a) A DE resonator in pre-stretched and actuated states[37]; (b) the natural frequency as a function of the applied voltage for a fixed $\lambda_{1p}$ and several values of $\lambda_{2p}$; (c) the natural frequency as a function of the applied voltage for a fixed $\lambda_{2p}$ and several values of $\lambda_{1p}$; (d) a two-layered DE resonator in two states[38]; (e) the resonance frequencies of five modes as a function of the pre-stretch ratio; (f) the resonance frequencies of five modes as a function of the radius of the passive layer. (Reproduced with permission from Li et al. [37], copyright 2012 by Elsevier; Reproduced with permission from Li et al. [38], copyright 2018 by IOP Publishing)

Apart from the above-mentioned works based on approximate analytical models, there also exist a series of theoretical studies on small-amplitude vibration responses of finitely deformed DE plates under mechanical and electric loads, in which the 3D theoretical framework summarized in Section 2 was employed. For example, for a pre-stretched electroactive rectangular plate described by an NH enriched DE material model [23], Wang et al. [40] investigated the free vibration behaviors under the application of an electric biasing field. The unique phenomena such as the softening effect of the increasing sizes and the stiffening effect of initial stretches on resonance frequencies were revealed by the numerical results. Therefore, an obvious competition effect on the frequencies was observed when the pre-stretch and electric biasing fields both play a role. Recently, based on the state-space formulations, Wang et al. [41] dev-eloped a general 3D theory to study the influence of pre-stretches and preset voltages on natural frequencies of compressible Gent rectangular DE plates of single-layered, two-layered, and even graded configurations. The stiffening effect of the pre-stretch and the softening effect of the applied voltage on frequencies as shown in **Fig. 5**(a) were observed. Moreover, they found that the gradient index, including the non-uniform material properties (see **Fig. 5**(b)), inhomogeneous pre-stretches (see **Fig. 5**(c)), and inhomogeneous electric voltages (see **Fig. 5**(d)), could be used to actively manipulate the natural frequencies of the layered and graded DE plates. Very recently, Cao et al. [42] derived the state-space formulations for the free vibration analysis of a pre-stretched DE circular plate, which is characterized by a generalized compressible NH model, under an external electric load. The same competition effect as the one found by Wang et al. [40,41] was also observed in the circular DE membrane. Moreover, they found that although a large biasing electric displacement helps to realize the thickness-stretch robustness, it may result in similar electromechanical instability to the ones revealed by Feng et al. [36] and Li et al. [37]. Consequently, they proposed that a thin circular DE plate could be an excellent choice to achieve thickness-stretch robustness and avoid electromechanical instability in frequency-tunable devices simultaneously.

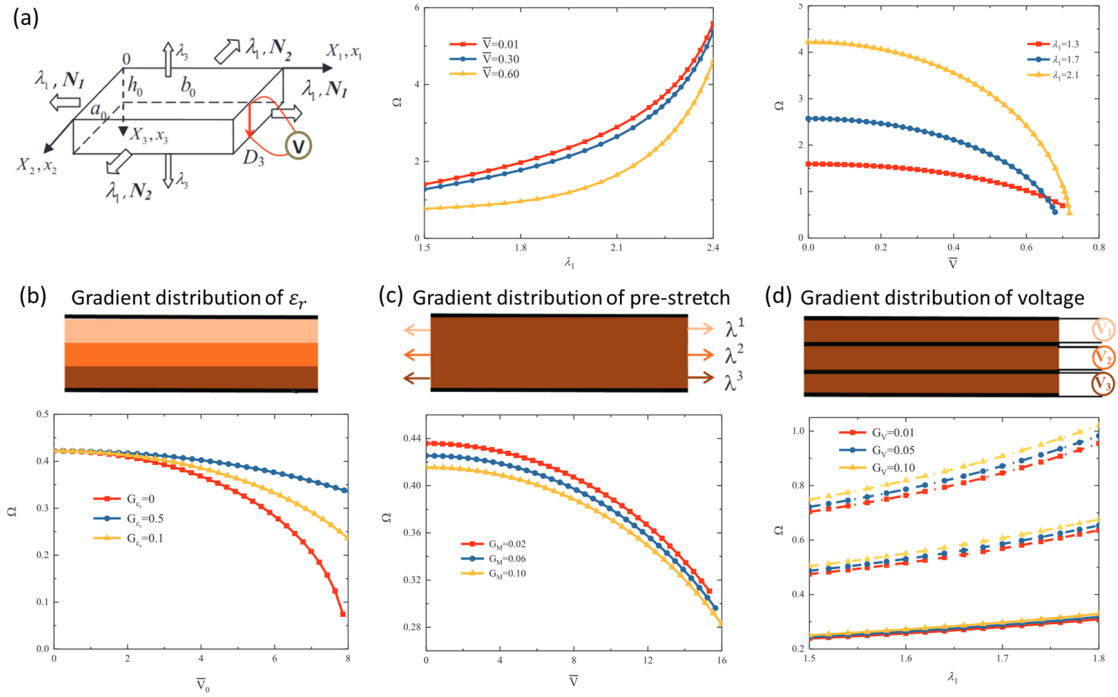

**Figure 5.** Flexural vibrations of pre-stretched DE single and graded plates under electric loads[41]: (a) Schematics of a DE single plate subjected to the combination of pre-stretch and voltage as well as the variations of the normalized frequency with pre-stretch for several voltages and with voltage for several pre-stretches; (b) variations of the normalized frequency with voltage for different gradient indexes of the permittivity; (c) variations of the normalized frequency with voltage for different gradient indexes of the pre-stretch; (d) variations of the normalized frequency with pre-stretch for different gradient indexes of the applied voltage. (Reproduced with permission from Wang et al. [41], copyright 2019 by Elsevier)

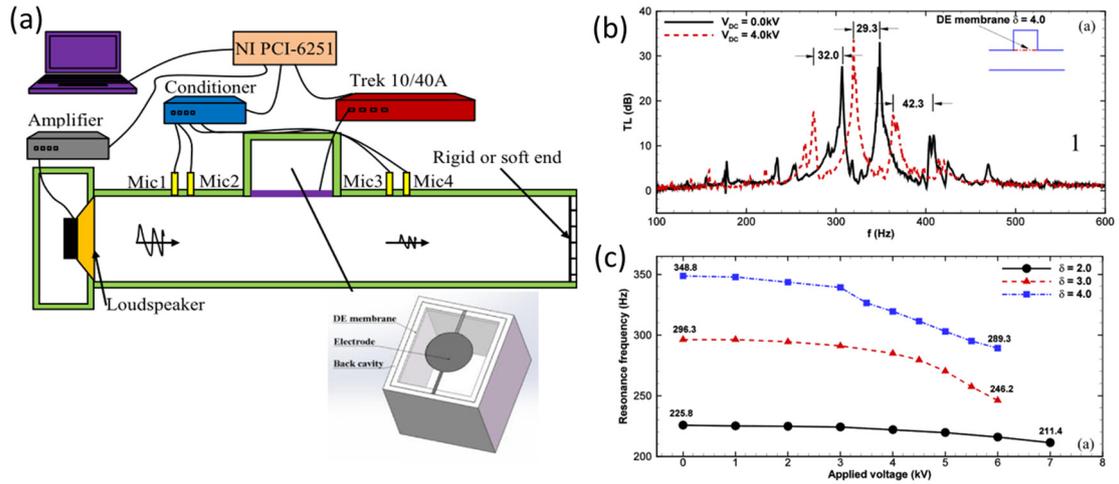

**Figure 6.** Novel tunable DE actuator for acoustic absorption[44]: (a) Acoustic measurement system of the DE duct silencer; (b) transmission loss under two applied voltages for the DE membrane with pre-stretch ratio $\delta = 4$; (c) resonance frequency of the duct silencer as a function of the applied voltage for different pre-stretch ratios. (Reproduced with permission from Lu et al. [44], copyright 2015 Acoustical Society of America)

In addition, some special mechanical components based on circular or rectangular DE membranes were proposed to achieve particular functions, e.g., noise reduction. For instance, Lu et al. [43] designed a frequency-tunable acoustic resonator formed by a circular DE membrane and an air back cavity by adjusting the applied voltage, which could be explored to fabricate a new generation acoustic absorber. The obtained results demonstrated that owing to the acoustic interactions between the DE membrane and the back cavity, more resonance peaks, at which the DE structure absorbs the acoustic energy efficiently, appear in the considered frequency range, and thus the resonator could suppress the noise at more frequencies. Moreover, the applied voltage lowers the resonance frequencies of the acoustic resonator and the resonance shift increases with the applied voltage, which provides an effective means to tune the frequency range of the acoustic absorber for noise reduction in real time. Subsequently, as shown in **Fig.**

**6**(a), Lu et al. [44] also designed a tunable acoustic resonator as a duct silencer, which is structurally formed by a pre-stretched DE membrane and a side-branch cavity. They found that both the pre-stretch ratio and applied voltage can be used to actively tune the resonance frequency and frequency range of the sound absorption (see **Fig. 6**(b) and (c)). Furthermore, based on the sub-structuring method, Yu et al. [45] developed a 3D analytical model to investigate the vibration properties of the same tunable NH DE resonator, which was subjected to the combination of the in-plane pre-stretch and external voltage, as the one designed by Lu et al. [44]. The transmission loss (TL) due to the acoustic coupling between the DE membrane and the side-branch cavity was calculated. The numerical results revealed that the locations and amplitudes of the peaks in the TL curves could be actively tuned by changing the pre-stretch ratio and applied voltage. Unlike the unique structures formed by the DE membrane and a cavity in Refs. [43,44,45], Jia et al. [46] directly placed a pre-stretched circular DE membrane inside the impedance measurement tube and measured the TL of the structure under different applied voltages. The results also demonstrated that the amplitudes and locations of peak frequencies depend on the initial stretch and the applied voltage.

Waves in solids significantly differ from their vibration responses. Numerous functional devices based on surface acoustic waves (SAWs) and bulk acoustic waves (BAWs) have been widely used in the fields of sensing, nondestructive testing, etc. In the last few decades, the development of soft DE materials enables researchers to manipulate the propagation characteristics of waves, which could be explored to design novel electrostatically tunable functional devices. Next, we summarize the advances in wave motions of soft DE plates of rectangular and circular configurations.

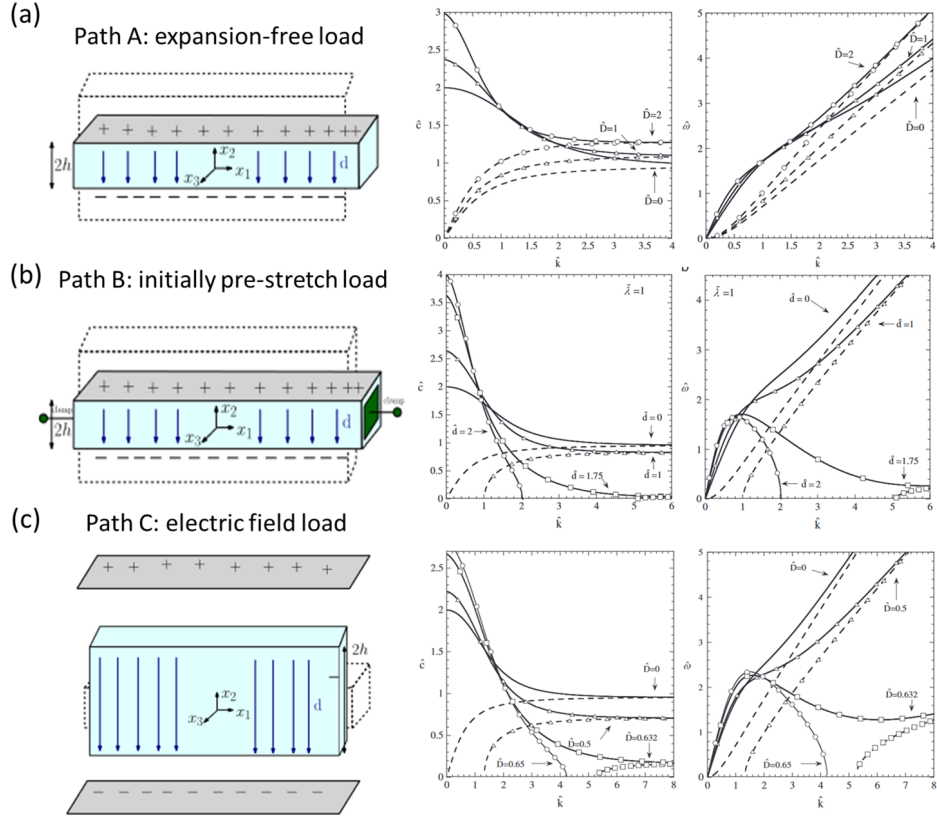

**Figure 7.** Rayleigh-Lamb waves propagating in DE layers subjected to large deformations induced by three loading paths[48]: (a) The DE layer under expansion-free load when subjected to the electric biasing field to soft electrodes and the corresponding dispersion relations of symmetric (continuous lines) and antisymmetric (dash lines) modes; (b) the DE layer under the combination of initial fixed pre-stretch and the electric biasing field to soft electrodes and the corresponding dispersion relations; (c) the DE layer immersed in a pre-existing electric biasing field and the corresponding dispersion relations. (Reproduced with permission from Shmuel et al. [48], copyright 2012 Elsevier)

Following the heuristic work by Dorfmann and Ogden [47] and based on the framework therein, a few theoretical works on small-amplitude waves propagating in soft DE plates subjected to pre-stretches and electric biasing fields have been carried out. Shmuel et al. [48] investigated the propagation of Rayleigh-Lamb waves in an in-plane pre-stretched soft DE rectangular plate subjected to an electric biasing field for three loading paths in **Fig. 7**. The corresponding dispersion relations for the finitely deformed DE layer described by an NH ideal DE material model were obtained. In path A, the DE layer coated with top and bottom soft electrodes was free to expand along the $x_1$ and $x_3$ directions. When the applied electric displacement increases, the wave velocities and frequencies of symmetric modes monotonously rise in the limit of short and long waves, while those of antisymmetric modes monotonously rise in the whole considered waves (see **Fig. 7**(a)). In path B, the DE layer coated with soft electrodes was initially pre-stretched along the $x_1$ direction and then clamped. With the increase of electric displacement field, the wave velocities of symmetric modes have a similar trend in the limit of long waves to those in path A, but have a reversed trend in the limit of short waves. With regard to the frequency, unlike the curves in path A, for some values of electric displacement, the curves are not monotonous. Interestingly, the threshold value of the wavenumber, beyond which the surface wave velocities and frequencies vanish, exists for certain values of electric displacement due to the occurrence of surface instability (see **Fig. 7**(b)). In path C, the DE layer was immersed in a pre-existing electric field. The increase in electric displacement has a similar influence on dispersion curves (see **Fig. 7**(c)) to those in path B. The threshold values of the wavenumber for certain electric displacements were also observed. In a word, the electric biasing field and pre-stretch significantly influence the wave velocity and frequency of the considered Rayleigh-Lamb waves, which could be explored to control the propagation velocity or to filter waves with certain wavelengths by applying proper electric biasing fields and pre-stretches. Furthermore, Galich and Rudykh [49] studied bulk waves (including longitudinal (P-) and transverse (S-) waves) propagating in finitely deformed DEs, which are characterized by compressible NH ideal or enriched DE material models in Refs. [32,50,51], under an electric load. They revealed that the influence of the electric field on wave propagation properties depends on the selected material models. Specifically, for the ideal DE model, the elastic waves are explicitly independent of the electric biasing field and only influenced by the deformation induced by the electric field. However, for the enriched DE model with all three electroelastic invariants taken into consideration, the elastic waves are directly affected by the electric biasing field. Moreover, the material compressibility results in the separation of the P- and S-waves due to the electric biasing field, indicating that the electric field could be used to manipulate the excitation of elastic waves in compressible DEs. Zhou and Chen [52] studied the influence of surface effect on surface waves propagating in a pre-stretched compressible NH DE half-space under an electric biasing field. They revealed that the surface effect enables the Love and generalized Rayleigh waves to become dispersive and size-dependent. Besides, the surface parameters, pre-stretch, and electric biasing field can change propagation velocities of surface waves, which provides an efficient method to actively tune wave characteristics of surface wave nanodevices. Recently, Broderick et al. [53] revisited the propagation of Lamb waves in soft DE plates subjected to mechanical and electrical loads. They obtained the explicit expressions for the dispersion relations in the cases of NH and Gent ideal dielectric models, and elucidated the effects of the electric biasing field, pre-stress and strain-stiffening on the wave characteristics. Based on the state-of-the-art finite-deformation viscoelasticity theory proposed by Hong [54] along with the theoretical framework in Section 2, Mohajer et al. [55] investigated the Rayleigh-Lamb waves propagating in a viscoelastic NH DE plate under the combination of pre-stretches and electric loads. Their numerical results revealed that the material viscoelasticity causes significant variations in wave propagation behaviors, e.g., the unique non-dispersive Rayleigh-Lamb wave for certain choices of material viscoelasticity and electromechanical loads. Moreover, they also found that both the mechanical and electric loads could be used to manipulate wave propagations in the DE plate. In addition to the theoretical research, Ziser and Shmuel [56] experimentally validated that the finitely deformed NH DEs can be used as tunable waveguides at low frequencies by investigating flexural waves in a pre-stretched DE film subjected to an external voltage. The experimental results showed that the applied voltage prominently reduces the measured wave velocity of the fundamental flexural wave mode, while increasing the initial stretch ratio enhances the wave velocity.

### 3.2 Solid and hollow cylinders

Different from the plate configuration, the cylindrical structure has some distinctive advantages [57], such as lower inactive-to-active material ratio, less bulky, and diverse deformation modes. Since Pelrine et al. [58] first proposed the soft DE tube actuator, the DE devices of the cylindrical configuration have attracted a great deal of attention from academia and industry. The relevant researches on small-amplitude vibrations and waves in finitely deformed soft DE cylinders of solid and hollow configurations have been performed in recent years, which are reviewed and summarized in this subsection.

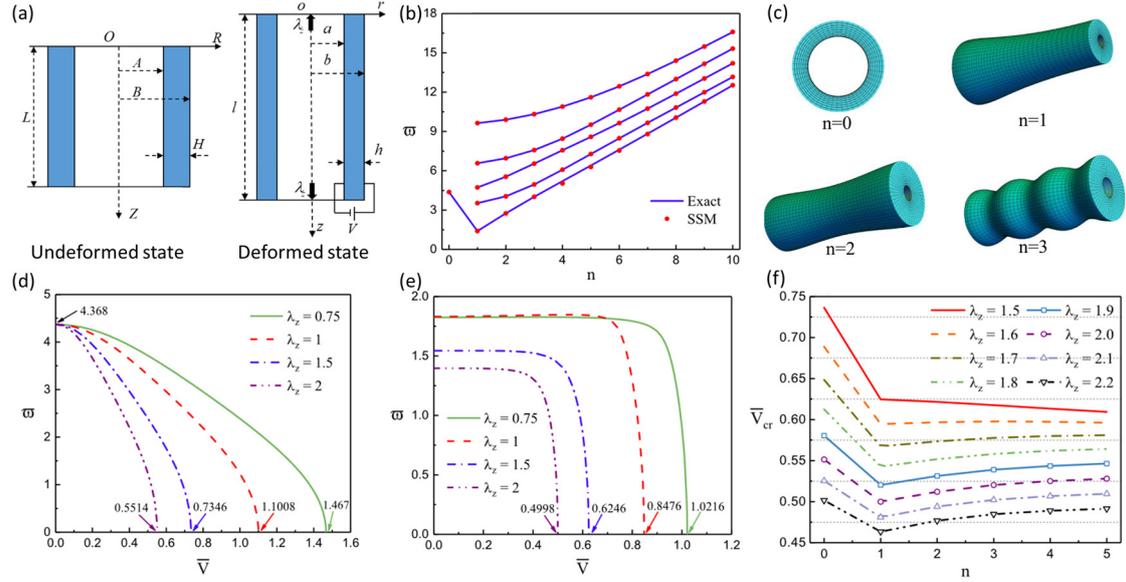

**Figure 8.** Electrostatically tunable axisymmetric vibration properties of a pre-stretched DE cylindrical tube under radial electric voltage[59]: (a) Schematics of a DE cylindrical tube in two states; (b) comparison of the first five resonance frequencies between the state-space method (SSM) and exact solutions versus the axial mode number n for the pre-stretched DE tube without applied voltage; (c) mode shapes of longitudinal vibrations for n=0 (breathing mode), 1, 2, and 3; (d) the first-order resonance frequency of the breathing mode n=0 as a function of radial electric voltage for the DE tube under different axial pre-stretches; (e) the first-order resonance frequency of the longitudinal vibration as a function of radial electric voltage for the DE tube under different axial pre-stretches with n=1; (f) critical voltages corresponding to the instabilities versus the axial mode number n for different axial pre-stretches. (Reproduced with permission from Zhu et al. [59], copyright 2020 Elsevier, Open Access)

With regard of vibrations, by employing the state-space method (SSM) with the approximate laminate model, Zhu et al. [59] investigated the axisymmetric torsional and longitudinal vibrations in an incompressible NH ideal DE cylindrical tube subjected to axial pre-stretch and inhomogeneous electric biasing fields induced by radial voltage as shown in **Fig. 8**(a). In absence of the applied voltage, the first five resonance frequencies of the pre-stretched DE tube obtained through the SSM agree well with those from the exact solutions for different axial mode numbers (see **Fig. 8**(b)). The mode shapes of the longitudinal vibrations were obtained and presented in **Fig. 8**(c). The relations between the first-order resonance frequencies and applied voltage for the pre-stretched DE tube with n=0 and 1 were presented in **Fig. 8**(d) and (e), and the critical voltage corresponding to the electromechanical or barreling instabilities, at which the resonance frequency of the relevant mode vanishes, was observed. Moreover, the numerical results revealed that the critical voltage depends remarkably on the initial pre-stretch and axial mode number (see **Fig. 8**(f)).

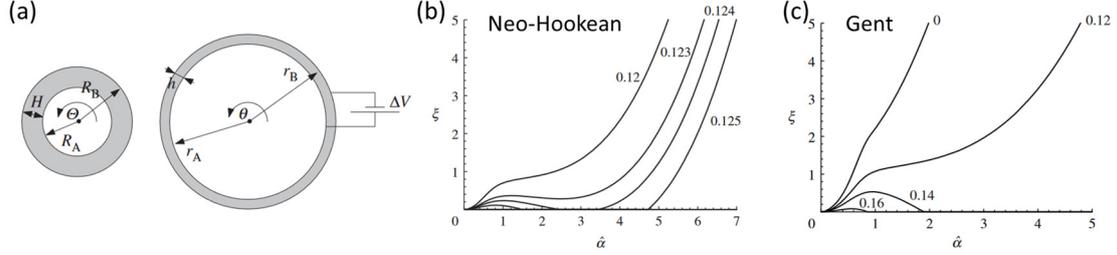

**Figure 9.** Axisymmetric waves in a finitely deformed DE circular cylindrical tube[62,63]: (a) Cross-section of a DE cylindrical tube in undeformed and deformed states[62]; (b) the dimensionless frequency $\xi = \rho\omega^2 A^2/\mu$ versus the dimensionless wavenumber $\hat{\alpha}$ under different electric voltages for an NH ideal DE tube[63]; (c) the dimensionless frequency $\xi = \rho\omega^2 A^2/\mu$ versus the dimensionless wavenumber $\hat{\alpha}$ under different electric voltages for a Gent ideal DE tube[63]. (Reproduced with permission from Shmuel and deBotton [62], copyright 2013 The Royal Society; Reproduced with permission from Dorfmann and Ogden [63], copyright 2020 The Royal Society)

More works on small-amplitude waves propagating in finitely deformed DE cylinders have been carried out. As early as 1980, Haughton [28] investigated symmetric and asymmetric waves in an incompressible elastic cylinder with finite rotating deformation about its axis. The numerical results indicated that symmetric wave modes such as pure torsional and longitudinal waves cannot propagate in the rotating cylinder. However, the wave velocity of asymmetric modes traveling up and down the cylinder diverges with the increase in angular velocity of rotation. Belward and Wright [29] theoretically and numerically calculated the 3D dispersion curves of small-amplitude waves in a prestressed cylinder characterized by the incompressible Mooney material. They revealed that the dispersion curves of the cylinder with initial deformation differ from that based on the classical infinitesimal elasticity theory, and the differences are dependent on the deformation manner, e.g., extension or compression along the axis of the cylinder. Based on the 3D linearized theory of elastic waves in initially stressed solids, Akbarov et al. [60] investigated the torsional waves propagating in a finitely pre-strained three-layered cylindrical tube. The results indicated that the initial strain has a notable influence on the wave velocity of the torsional waves, which is also dependent on the mode order and wavenumber.

Much effort has been devoted to theoretically investigating small-amplitude waves in finitely deformed soft DE cylinders under mechanical and electric biasing fields through the 3D theoretical framework summarized in Section 2. For instance, Chen and Dai [61] investigated the axisymmetric waves propagating in an incompressible NH ideal DE solid cylinder subjected to homogeneous pre-stretches and axial electric displacement. The numerical results showed that the dispersion curves of the axisymmetric waves in such a DE cylinder have a significant correlation with the pre-stretch and electric biasing field. Shmuel and deBotton [62] also analyzed the longitudinal axisymmetric waves in a pre-stretched incompressible NH ideal DE cylindrical tube, but subjected to the radial electric biasing field, and the influences of tube geometry, pre-stretch, and electric biasing field on the propagation behaviors (wave velocity and frequency vs wavenumber) of the fundamental mode were discussed. The results demonstrated the feasibility of manipulating longitudinal waves propagating in a soft DE cylindrical tube by controlling the mechanical and electric loads. Moreover, a threshold value of the electric biasing field (when applying a larger electric biasing field than the threshold value, there exists a cut-off wavenumber beyond which the wave velocity vanishes) was observed, and the threshold value is significantly dependent on the tube geometry and pre-stretch, which provides a potential strategy to filter waves with certain wavelengths of interests. Recently, Dorfmann and Ogden [63] restudied the issue considered by Shmuel and deBotton [62], and some different results, particularly related to the cut-off wavenumber, were obtained as shown in **Fig. 9**. Specifically, they revealed that for a larger voltage than the threshold value, there exists a finite cut-off wavenumber band (see **Fig. 9**(b)), within which the frequency vanishes, for the incompressible NH ideal dielectric model. Moreover, Dorfmann and Ogden found that unlike the incompressible NH model, there are no real frequencies (i.e., no propagating waves) when the wavenumber exceeds its electric-voltage-dependent cut-off value for the Gent material model (see **Fig. 9**(c)). Furthermore, Shmuel [64] studied the tunability of torsional waves propagating in a pre-stretched DE cylindrical tube by adjusting a radial electric voltage, and systematically analyzed the dependency of propagation characteristics on tube thickness, pre-stretch, and applied voltage. It was found that the wave velocity and frequency of the torsional waves are remarkably influenced by these factors. Besides, the axial initial tension leads to a limit value of the applied voltage (beyond which the higher-order torsional modes disappear in the considered wavenumber range due to the collapse of the DE cylindrical tube), and the threshold voltage decreases with the increase in the initial tension. Different from the axisymmetric waves in DE solid or hollow cylinders mentioned above, the propagation of non-axisymmetric waves in a finitely deformed soft electroactive hollow cylinder under axial electric biasing fields was studied by Su et al. [65] in detail. The stiffening effect of the electric biasing field, which results in the increase in phase velocity and frequency of non-axisymmetric waves in the hollow cylinder, was revealed. They also declared that although increasing the axial pre-stretch raises the frequency and phase velocity in the axially infinite cylinder, the pre-stretch does not always increase the frequency (and the overall material rigidity) of a finite cylinder due to the competition effect between the increased size and initial stretch, which has been revealed by Wang et al. [40,41]. In addition, by using the SSM along with the approximate laminate model, Wu et al. [66] studied both the guided circumferential shear-horizontal (SH-type) and Lamb-type waves in an incompressible NH ideal DE cylindrical tube subjected to inhomogeneous biasing fields in detail (see **Fig. 10**(a) and (b)). The numerical results in **Fig. 10**(c) demonstrated the efficiency and accuracy of the SSM for solving the dispersion relations. It was found that the propagation characteristics of these two types of guided circumferential waves are significantly dependent on the geometric parameters and inhomogeneous biasing fields (see **Fig. 10**(d) and (e)), and interesting phenomena such as the frequency veering in **Fig. 10**(f) were revealed by the numerical examples. Subsequently, based on their proposed SSM method, Wu et al. [67,68] additionally investigated the axisymmetric torsional and longitudinal waves propagating in a functionally graded (FG) DE hollow cylinder under the combined loads of axial pre-stretch, radial pressure difference, and radial electric voltage. They used the Mooney-Rivlin ideal dielectric material model with the material constants varying along the radial direction in an affine way to characterize the incompressible isotropic DE tube. The numerical results illustrated that the pre-stretch, radial pressure difference, and radial voltage can be utilized to steer the propagation properties (e.g., frequency and wave velocity) of the guided axisymmetric waves in the FG DE tube, and the tunable capacity can be significantly enhanced by increasing the material gradient.

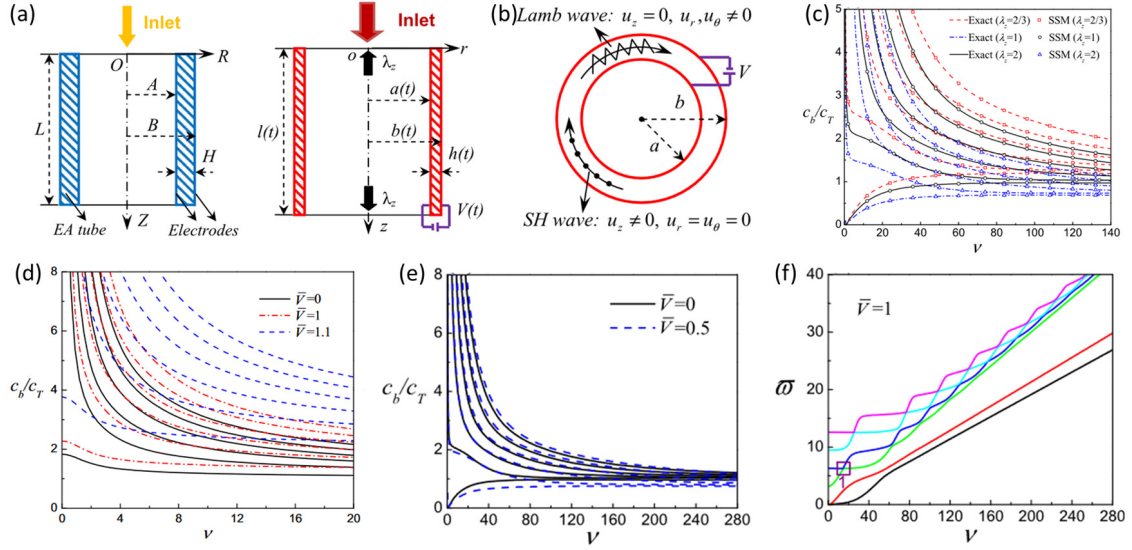

**Figure 10**. Guided circumferential SH-type and Lamb-type waves in soft DE cylindrical tubes subjected to radially inhomogeneous biasing fields[66]: (a) A DE tube in undeformed and actuated states; (b) two types of incremental guided circumferential waves superimposed on the actuated state; (c) comparison of phase velocity spectra for the first six Lamb-type wave modes obtained by the SSM and the exact solutions for the tubes at different pre-stretches without applied voltage; (d) phase velocity spectra of SH-type waves for a DE tube under different applied voltages; (e) phase velocity spectra of Lamb-type waves for a DE tube under different applied voltages; (f) unique frequency veering phenomenon in the rectangle labeled with 1. (Reproduced with permission from Wu et al. [66], copyright 2016 Elsevier)

### 3.3 Hollow spheres

The soft DEs of spherical shape have potential applications as pumps, loudspeakers, elements of shell-like actuators, etc. Compared with the works about soft DEs of plate or cylindrical configurations, the investigations on small-amplitude vibrations and waves in soft DE spheres are very limited. Zhu et al. [69] investigated the small-amplitude oscillations around the finitely deformed NH balloon subjected to a combination of pressure and voltage. The numerical results indicated that applying a constant pressure or electric voltage can tune the resonance frequency of the DE spherical balloon. Besides, the safe range of the pressure and voltage should be controlled to avoid the destruction of the soft DE balloon. Based on the SMM along with the approximate laminate technique proposed by Wu et al. [66], Mao et al. [70] investigated the torsional and spheroidal vibrations in an incompressible NH and Gent ideal DE balloon subjected to inhomogeneous biasing fields induced by the combination of radial electric voltage and internal pressure as shown in **Fig. 11**(a). It was found that the SSM is also suitable for the vibration analysis of the DE spherical balloon under an inhomogeneous biasing field with high precision. The influence of snap-through instability on vibration behaviors in soft DE structure was first discussed. The obtained results demonstrated that the resonance frequencies of various vibration modes in the DE balloon can be tuned flexibly by adjusting the radial electric voltage and internal pressure (see **Fig. 11**(b)). More interestingly, the snap-through instability induced by the electromechanical biasing fields for the Gent DE balloon (see **Fig. 11**(c)) can realize the sharp shift of the resonance frequency when the applied voltage exceeds the critical value as shown in **Fig. 11**(d).

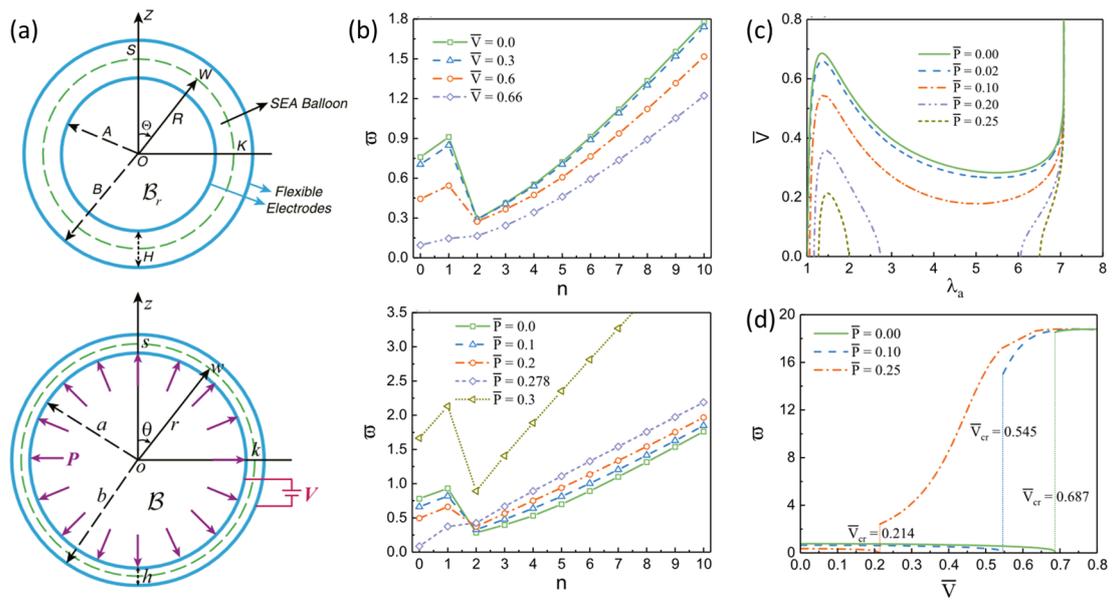

**Figure 11.** Small-amplitude vibration responses of pressurized DE spherical balloons subjected to radially applied electric voltage[70]: (a) Cross-section of a DE balloon in the undeformed and deformed states; (b) resonance frequency versus the angular mode number n for a Gent DE spherical balloon under different voltages and internal pressures; (c) Variation of the dimensionless electric voltage with the stretch ratio for different internal pressures; (d) resonance frequency of spheroidal vibrations as a function of the dimensionless electric voltage for the breathing mode (n=0) under different internal pressures. (Reproduced with permission from Mao et al. [70], copyright 2019 Elsevier)

**Table 1. Summary of small-amplitude vibrations and waves in finitely deformed DE structures of the plate, cylindrical, and spherical configurations. In the ideal DE model, only the electroelastic invariant $I_5$ is considered[32], while in the enriched DE model, in addition to $I_5$, the electroelastic invariants $I_4$ or both $I_4$ and $I_6$ are considered[23,47,71]. Acronyms: DE (dielectric elastomer), NH (neo-Hookean), FG (functionally graded).**

| Configurations | Dynamic responses | Material models | References |
| --- | --- | --- | --- |
| Plate | Vibration | \ | [30,36,43,44] |
| Plate | Vibration | NH ideal DE model | [31,38,45] |
| Plate | Vibration | Gent ideal DE model | [37,41,46] |
| Plate | Vibration | Visco-hyperelastic ideal DE model[34] | [33] |
| Plate | Vibration | Mooney-Rivlin ideal DE model | [35] |
| Plate | Vibration | Yeoh ideal DE model | [39] |
| Plate | Vibration | NH enriched DE model | [40,42] |
| Plate | Wave | Viscoelastic NH ideal DE model | [55] |
| Plate | Wave | NH ideal DE models | [48,52,56] |
| Plate | Wave | NH ideal or enriched DE models | [49] |
| Plate | Wave | NH and Gent ideal DE models | [53] |
| Cylinder | Vibration | NH ideal DE model | [59] |
| Cylinder | Wave | NH and Gent ideal DE models | [63] |
| Cylinder | Wave | NH ideal DE model | [62,64,66] |
| Cylinder | Wave | NH enriched DE model | [61,65] |
| Cylinder | Wave | FG Mooney-Rivlin ideal DE model | [67,68] |
| Sphere | Vibration | NH and Gent ideal DE models | [69,70] |

## 4. Nonlinear dynamic responses of DE structures

When subjected to a time-dependent load, the large amplitude and high-frequency excitation may result in nonlinear dynamic responses of the DE structures, which have attracted intensive research attention in many areas of science and engineering. Focusing on the main topics of this article, the recent advances in nonlinear vibrations and waves in soft DE structures are reviewed and summarized in this section. Furthermore, since the electromechanical instability frequently accompanies the nonlinear vibrations, the instability phenomena will be also referred to more or less.

## 4.1 Nonlinear vibrations

Nonlinear vibrations in DE structures are a classical and significant topic due to the wide applications of DEs in loudspeakers, active noise control, frequency tuning, energy harvesting, etc. Many research groups have made great effort on this topic, and a lot of meritorious works have been reported in recent years.

### 4.1.1 Rectangular and circular plates

Zhu et al. [31] explored the nonlinear vibrations in a pre-stretched DE circular membrane under the combination of the out-of-plane pressure and external voltage. They found that both the sinusoidal mechanical and electric loads excited the superharmonic (whose frequency is about double that of excitation), harmonic (whose frequency is close to that of excitation), and subharmonic (whose frequency is about half of that of excitation) resonances, which agreed well with the experimental observations. Similar nonlinear oscillations, which resonate at multiple frequencies of excitation in a pre-stretched DE plate under the sinusoidal electric voltage, were also observed by Li et al. [37]. In addition, Dai and Wang [72] investigated the nonlinear oscillations of a finitely deformed complex system composed of an NH DE rectangular membrane, springs, and a pair of slide tracks. The numerical results showed that for free vibration the system exhibits periodic oscillation with the amplitude significantly influenced by the initial stretch, while the connected linear/nonlinear springs could change the static equilibrium position of the system. Moreover, when applying a harmonic excitation, the system undergoes quasi-periodic oscillations and resonates with multiple frequencies of excitation, also resulting in superharmonic and subharmonic resonances. Subsequently, replacing the mechanical load via spring components in Ref. [72] with the electric load, Dai et al. [73] studied the dynamic responses of a DE rectangular membrane sandwiched by compliant electrodes when subjected to electric loads. They found that when applying a constant electric load, the DE membrane undergoes finite-amplitude periodic oscillation. However, when subjected to harmonic electric excitation, the quasi-periodic, chaotic, and chaotic divergent oscillations were observed, indicating the transition of the system from stability to instability. Furthermore, based on the Mooney-Rivlin and Ogden material models, Kim et al. [74] focused on calculating the dynamic instability parameters in the linear and nonlinear vibrations of a pre-stretched DE rectangular membrane subjected to a step voltage. The obtained critical values of the dynamic instability parameters can be used to judge the transition from periodic oscillations to nonperiodic ones of the system. Recently, free and forced nonlinear vibrations with large amplitude in a sandwiched microbeam resonator composed of the Yeoh DE rectangular film were studied by Ariana and Mohammadi [75]. The critical voltage (below which the resonator exhibits a stable periodic oscillation) and saddle-node bifurcation were found to characterize the transition from stability to instability. Moreover, the forced excitation amplitude, constant voltage, detuning parameter, and beam aspect ratio have significant influences on the saddle-node bifurcation. Cooley and Lowe [76] reported the extreme stretchability of a DE circular membrane by applying the sinusoidal voltage. They revealed that the upper branches with large-amplitude stretch occur at the jump frequencies, where small variations in frequency may yield abrupt changes in the dynamic stretch, and these branches can also be achieved by superimposing the impulse voltage on the sinusoidal one with the excitation frequency unchanged.

Much effort from researchers has been made on the study of the damping effect of material viscoelasticity on nonlinear vibration responses in DE membranes. For instance, based on the Euler-Lagrange equation, Xu et al. [77] proposed an analytical model to analyze the vibration behaviors of an incompressible NH DE rectangular membrane subjected to an electric biasing field. With the increase in the applied voltage, the transition phenomenon from stability to instability was observed in the frequency range of excitation. Moreover, due to the damping effect, the original steady oscillation becomes a constant vibration, and the increase in the damping coefficient enhances the first-order resonance frequency. With the increase in the pressure or voltage, the stability transition was also revealed by Wang et al. [78] by investigating the viscoelastic effect on the nonlinear vibrations in an NH DE circular membrane, whose viscoelastic behavior is characterized as a rheological model of springs and dashpots. Zhou et al. [79] examined the dynamic responses of a tunable viscoelastic DE membrane oscillator under static and harmonic external voltage. They found that the resonance frequency of the DE oscillation can be significantly tuned by applying a static external voltage, while the oscillation patterns of interest can be achieved by adjusting both the static and alternating voltages. Liu et al. [80] experimentally studied the nonli-

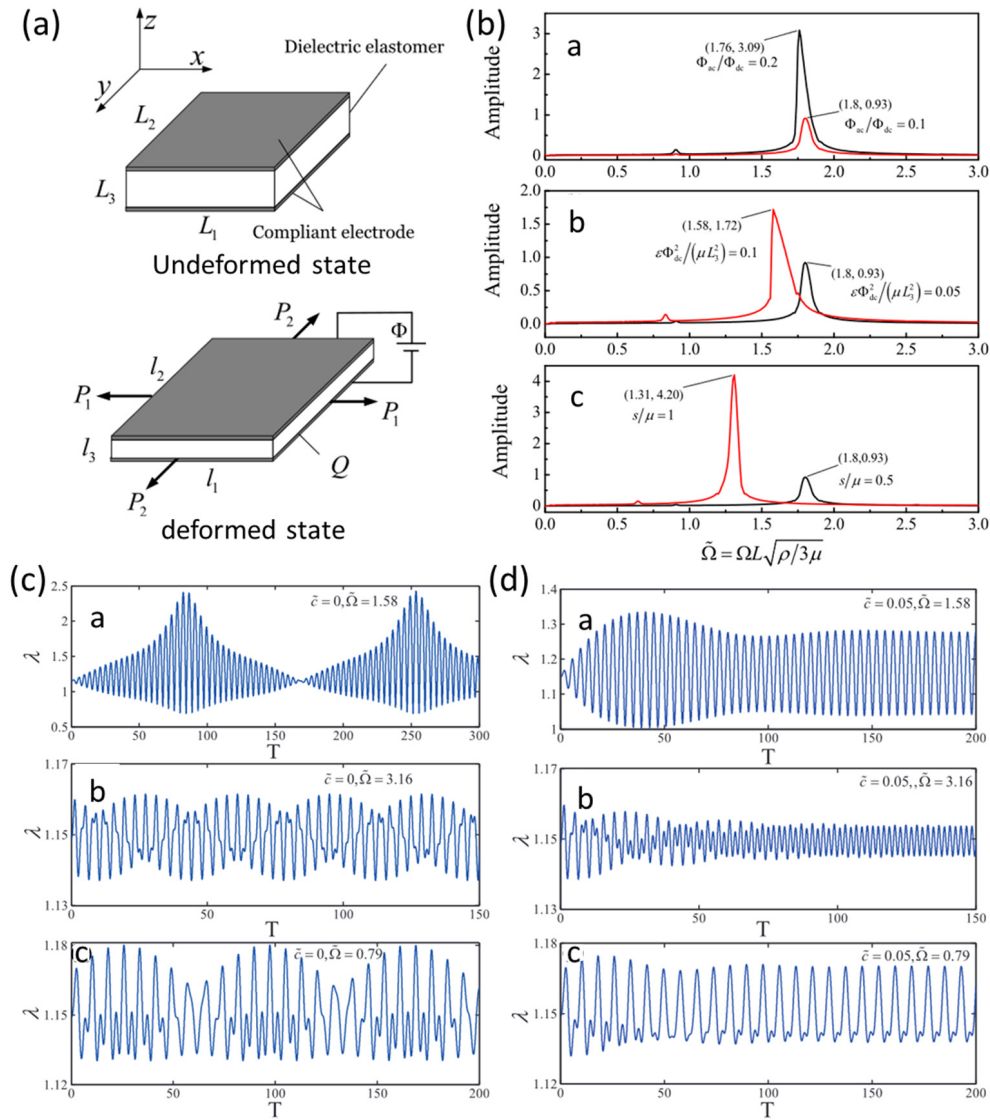

**Figure 12.** The influence of damping effect on nonlinear dynamic responses of a DE rectangular membrane subjected to the combination of mechanical force and electric voltage[81]: (a) A DE membrane in two states; (b) the oscillation amplitude as a function of the dimensionless excitation frequency for different AC voltages, DC voltages, and pre-stresses; (c) oscillation responses without damping effect for different excitation frequencies; (d) oscillation responses with damping effect for different excitation frequencies. (Reproduced with permission from Sheng et al. [81], copyright 2014 IOP Publishing)

near dynamic responses of an in-plane finitely deformed incompressible DE circular membrane under alternating electric voltage and discussed the effect of viscoelasticity, peak voltage, frequency of excitation, and waveform on the deformation equilibrium position of the DE membrane. The obtained results showed that the high viscoelasticity and peak voltage result in a severe drift of the equilibrium position during the vibration of the membrane, while the large pre-stretch could reduce the drift, which may be explored to realize large oscillation displacement with small drift by balancing several factors. Based on a free-energy model, Sheng et al. [81] theoretically analyzed nonlinear dynamic responses of an in-plane deformed incompressible and viscoelastic DE rectangular membrane subjected to the combination of mechanical force and electric voltage as shown in **Fig. 12**(a). They found that the applied force and electric voltage could be used to actively tune the natural frequencies and oscillation amplitudes of the system (see **Fig. 12**(b)). Moreover, when applying a sinusoidal voltage, the DE membrane without the damping effect undergoes a nonlinear quasi-periodic oscillation (see **Fig. 12**(c)), while the introduction of the damping effect may turn the quasi-periodic oscillation into a constant vibration with a decreasing vibration amplitude (see **Fig. 12**(d)). Subsequently, the same research group [82] also investigated the nonlinear vibrations of a dissipative system formed by an in-plane deformed DE rectangular membrane under alternating mechanical load and focused on the influences of two dissipative processes, i.e., viscoelasticity and current leakage, on the oscillation of the DE membrane. The numerical results indicated that for a sinusoidal mechanical load and static voltage, the system resonates at several excitation frequencies. The material viscoelasticity reduces the natural frequency of the system but increases the mean stretch of oscillation. However, the current leakage after cutting off the power source has the opposite effect to the viscoelasticity. Besides, they developed a dynamic analytical model to study nonlinear vibrations in a viscoelastic DE membrane subjected to the combination of non-equibiaxial tensile force and external voltage [83]. It was found that when two in-plane tensile forces are different, the amplitude-frequency responses exhibit strong nonlinearity, but when they are equal to each other, the system vibrates less nonlinearly and exhibits the beating phenomenon. In addition, based on the squeeze-film damping model, Feng et al. [84,85] utilized the parameterized perturbation method to investigate the influence of ambient viscosity on the nonlinear dynamic responses of a pre-stretched double-clamped DE beam resonator under the time-dependent external load. It was observed that the pressure of the squeeze film, applied voltage, and initial stretch can prominently alter the resonance frequency and quality factor of the DE beam-based resonator. Barforooshi and Mohammadi [86]

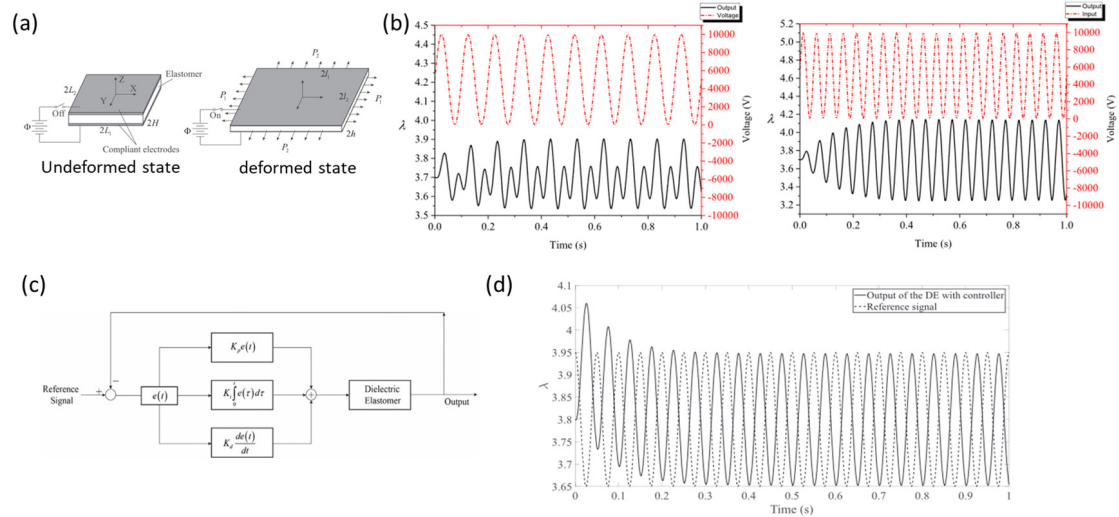

**Figure 13.** Nonlinear vibration responses and active control of a visco-hyperelastic DE rectangular membrane[87]: (a) A DE membrane in two states; (b) Time-domain responses of the DE membrane under super-harmonic frequency and harmonic frequency; (c) block diagram of the closed-loop feedback control system of the DE membrane with a PID controller; (d) Time-domain response of the DE membrane with the controller. (Reproduced with permission from Li et al. [87], copyright 2018 Elsevier)

also investigated the nonlinear vibrations of the DE microbeam resonator similar to the ones in Refs. [84,85] with the geometric (von Kármán strain-displacement relation) and the material (Yeoh model) nonlinearities taken into consideration. The numerical results indicated that with the increase in mode number, the influences of structural aspect ratio and initial maximum amplitude on nonlinear frequency become more prominent. Li et al. [87] studied the influence of viscoelasticity and strain-stiffening on the resonance frequency and oscillation amplitude of a Gent DE rectangular membrane under alternating electric voltage as shown in **Fig. 13**(a). It was found that due to the viscoelasticity and nonlinear oscillation, the output oscillations (e.g., their periodicity and amplitude) significantly deviate from the desired harmonic oscillations (see **Fig. 13**(b)). Then the authors designed a closed-loop feedback active control system in **Fig. 13**(c) to eliminate the influence of nonlinear effect, beat phenomenon, and phase lag, and finally realized the desired output (see **Fig. 13**(d)). Based on the Zener rheological model consisting of springs and a damper, Khurana et al. [88] discussed the effect of material viscoelasticity on the dynamic responses of an NH DE minimum energy structure under electric loads. They found that the increase in viscosity parameter could delay the onset of the finite-amplitude equilibrium oscillations of the system under a constant electric voltage. When applying an alternating voltage, the stability transition (stable-unstable-stable) was observed from the Poincaré maps and phase diagrams, indicating that the system undergoes a transition from quasiperiodic to aperiodic oscillation. With the material and geometric nonlinearities taken into consideration, Alibakhshi and Heidari [89] investigated the nonlinear vibrations of a Gent DE microbeam resonator under sinusoidal voltage and discussed the chaos phenomenon and damping effect therein. The numerical results showed that for the sinusoidal voltage excitation, the resonator without damping undergoes the nonlinear quasi-periodic oscillation, while the resonator with damping effect goes through a nonlinear non-periodic attenuation vibration and then evolves to a constant periodic oscillation, namely the chaotic vibration. Next, Alibakhshi et al. [90] studied the nonlinear free and forced vibrations of a Cosserat DE microcantilever, as a novel atomic force microscopy, when the static and alternating voltages were applied and at the same time the size effect and damping effect were taken into account. Similarly, they found that the damping effect lowers the response amplitude of the system and the sinusoidal voltage results in nonlinear quasi-periodic vibrations. Besides, the increase in the size-effect parameter enhances the nonlinear frequency and the response amplitude for the undamped system. However, increasing the size effect lowers the response amplitude of the damped system. In addition to linear viscoelasticity, Li et al. [91] used the framework of finite-deformation viscoelasticity to analyze the influence of nonlinear material viscosity on the dynamic response of a DE rectangular membrane subjected to an electric field. It was found that the nonlinear viscosity only affects the transient responses of the frequency and stretch ratio during the frequency tuning process, but its influence on the steady frequency and stretch ratio after sufficient time is neglectable. Zhang et al. [92] designed a spring roll actuator based on a viscoelastic DE membrane and investigated the static and dynamic responses of the system under static and time-dependent electric loads, respectively. It was found that there exists a critical static voltage for the DE actuator, below which the deformed system could reach equilibrium after the viscoelastic relaxation, but beyond which no equilibrium oscillation can be reached. When applying a time-dependent voltage, the frequency of excitation, material viscoelasticity, and spring stiffness can control the oscillation amplitude and natural frequency of the DE actuator to avoid the resonance so as to get rid of its failure. In quick succession, Zhang et al. [93] designed a tunable active vibration damper composed of a viscoelastic DE membrane and a spring oscillator by adjusting the applied voltage, initial stretch, and excitation frequency. The results showed that the passive vibration control is realized by changing the material viscoelasticity and spring stiffness, while the active vibration attenuation under an opposite-phase sinusoidal voltage could be achieved by modulating the excitation frequency and voltage amplitude.

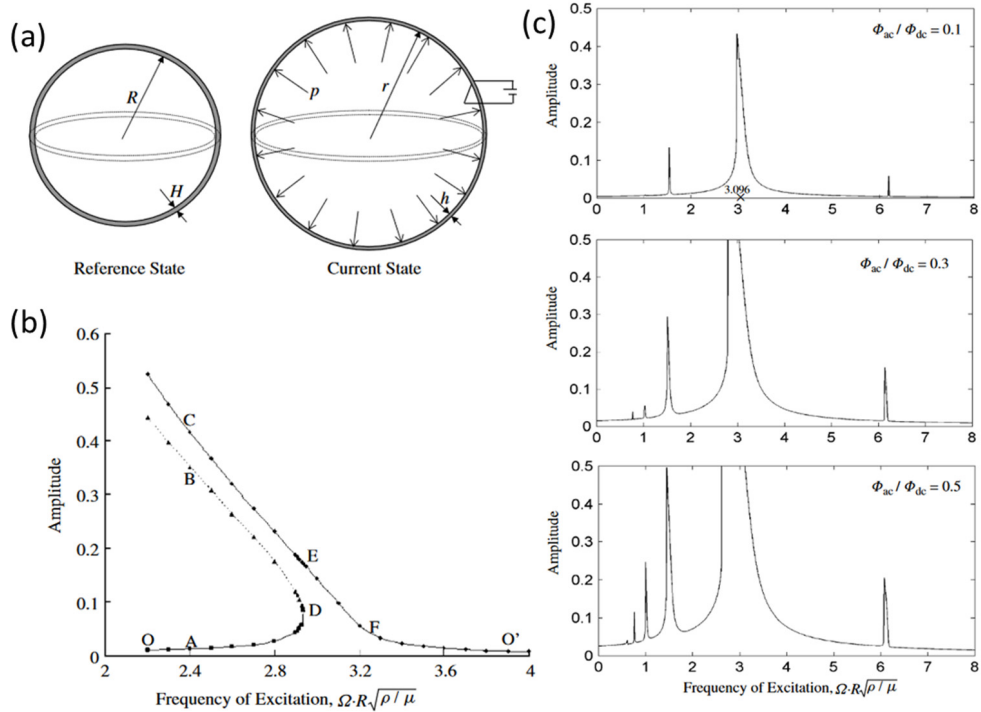

**Figure 14.** Nonlinear oscillation of a DE spherical balloon subjected to the combination of internal pressure and radially applied electric voltage[69]: (a) A DE spherical balloon in two states; (b) steady-state amplitude versus the excitation frequency; (c) the oscillation amplitude as a function of the excitation frequency for several values of $\Phi_{ac}/\Phi_{dc}$ when applying a sinusoidal voltage $\Phi_{dc}+\Phi_{ac}\sin(\Omega t)$. (Reproduced with permission from Zhu et al. [69], copyright 2010 John Wiley and Sons)

One recent interesting work about nonlinear vibrations accomplished by Wang et al. [94,95] was finally noticed in this subsection. Based on the Timoshenko beam theory, they [94] investigated the nonlinear free vibrations of graphene platelets (GPLs) reinforced DE composite beams subjected to an electric biasing field. The results indicated that the nonlinear frequency notably depends on the GPL concentration, GPL diameter-to-thickness ratio, and the applied electric field. A transition region, within which the slight variation of the excitation frequency could cause a dramatic change in the nonlinear frequency, was revealed. Subsequently, built upon the Timoshenko beam theory and von Kármán geometric relation, the nonlinear static bending and dynamic vibrations of GPL-reinforced DE beam subjected to the electric biasing field and external mechanical load were studied by Wang et al. [95]. The influences of GPL concentration, GPL aspect ratio, and applied electric field on the static and forced vibration were discussed in detail and a similar transition region was also observed. Besides, the increase in the GPL concentration, due to its enhanced effect on the dielectric property, raises the deflection of the beam when applying the electric biasing field.

### 4.1.2 Hollow cylinders and spheres

There also exist some researches on nonlinear vibrations in soft DE structures of hollow cylinders and spheres. Zhu et al. [69] theoretically investigated the nonlinear vibrations of a pre-stretched DE spherical balloon subjected to the combination of internal pressure and radially applied electric voltage as shown in **Fig. 14**(a). The numerical results showed that when applying a static pressure and a sinusoidal voltage, the spherical balloon resonates at multiple values of the excitation frequency as shown in **Fig. 14**(c), resulting in similar superharmonic, harmonic, and subharmonic modes to the ones in Ref. [31]. Moreover, similar to Refs. [96,97], the hysteresis phenomenon was revealed that the oscillating amplitude of the steady-state balloon will jump to certain values (like from point O to A, to D, then jump to point E, to F, and to O' as in **Fig. 14**(b)) when the excitation frequency increases continuously. Different from the thin spherical shell model in Ref. [69], Yong et al. [98] studied the nonlinear vibrations of a pre-stretched thick-walled DE spherical shell governed by the incompressible NH model under electric loads, and discussed the stability of the system when applying the sinusoidal voltage. It was found that the DE spherical shell can undergo a nonlinear quasi-periodic oscillation when the voltage is within the critical value, and the increase in the shell thickness helps to stabilize the system. However, an electric voltage beyond the critical value may destroy the spherical shell. Furthermore, Bortot [99] analyzed the nonlinear vibrations of a thick-walled DE cylindrical tube under electric loads. The simulation results revealed that for a constant electric field, the resonance frequency and the oscillation amplitude can be significantly influenced by the tube thickness. For a harmonic electric load, the DE tube can undergo the superharmonic resonance and also resonate near the frequency of excitation, which is characterized by the beat phenomenon. Son and Goulbourne [100] developed a numerical model to investigate the dynamic actuation and sensing response of tubular DE transducers. Good agreement with 3% overall error between the maximum values of numerical and experimental results demonstrated that the dynamic model combined with the solution approach based on the finite difference method provides an effective means for predicting dynamic responses of tubular DE transducers.

### 4.2 Nonlinear waves

Next, we pay attention to the nonlinear waves propagating in DE structures. The relevant studies remain limited and most of them focus on nonlinear waves in elastic materials/structures, which include weakly nonlinear waves (small but finite amplitude) and strong nonlinear waves (no assumption of a small wave amplitude). Some early works have reported weakly nonlinear waves in elastic rods. For example, Porubov and Samsonov [101] developed a refined model for the propagation of longitudinal strain waves in a rod with nonlinear elasticity and provided the solitary wave solutions of the refined model. Cohen and Dai [102] derived the model equations for weakly nonlinear waves propagating in a compressible Mooney-Rivlin rod and demonstrated the existence of far-field solitary waves. Dai and Fan [103] also found the far-field solitary waves in a cylindrical elastic rod composed of a compressible Murnaghan material. Recently, Dai and Peng [104] studied weakly nonlinear long waves in a pre-stretched cylinder of a Blatz-Ko material by employing the method of coupled series and asymptotic expansions. The numerical results indicated that the geometric softening induced by the uniaxial tension leads to the appearance of three types of nonlinear waves (solitary, periodic, and kink waves). Moreover, the excitations of different nonlinear waves can be tuned by changing the axial pre-stretch. Furthermore, the kink and kink-like waves propagating in pre-stretched Mooney-Rivlin viscoelastic rods were investigated by Wang et al. [105]. The numerical examples indicated that the pre-stretch and viscosity influence the wave shape and nonlinear wave velocity. Therefore, the pre-stretch could be used to modulate the kink and kink-like waves and these two types of waves may be utilized to measure the viscosity of the material. Destrade et al. [106] studied weakly nonlinear elastic shear waves composed of an anti-plane shear motion and a general in-plane motion in isotropic incompressible nonlinear elastic solids and found that the anti-plane and in-plane motions cannot be decoupled and that linear polarization is impossible for general nonlinear two-dimensional (2D) shear waves. Aiming at the soft DEs, Wang et al. [107] investigated the propagation characteristics of solitary waves in an incompressible NH DE cylindrical rod subjected to an electric biasing field as shown in **Fig. 15**(a). It was found that the wave velocity (see **Fig. 15**(b)) and wave shape (see **Fig. 15**(c)) of the solitary waves can be tuned by the biasing electric displacement, which could help to experimentally excite the solitary waves with certain features of interest by using different electric biasing fields. In addition, solitary waves propagating in a hard magneto-electro-elastic (MEE) circular rod were studied by Xue et al. [108]. The numerical examples not only demonstrated the existence of the solitary waves in MEE materials but also illustrated that the multi-field coupling has a significant influence on nonlinear wave characteristics.

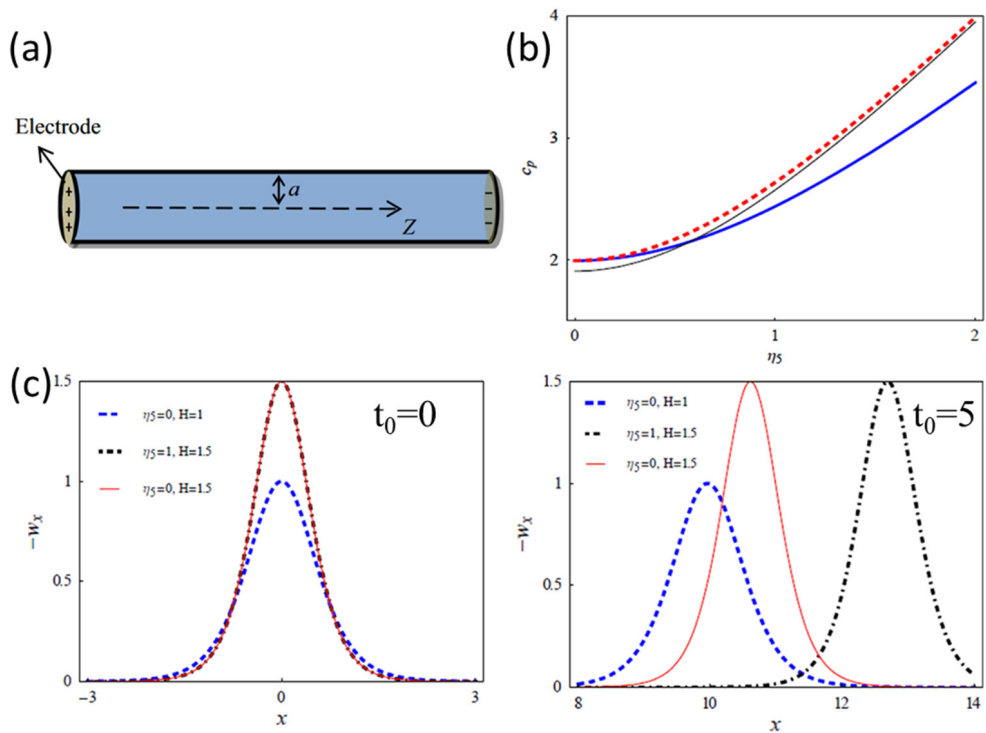

**Figure 15.** Propagation of adjustable solitary waves in DE cylindrical rods[107]: (a) A DE rod subjected to an axial electric biasing field applied through the end electrodes; (b) influence of the dimensionless biasing electric displacement $\eta_5$ on the velocity of solitary waves for different DE material parameters; (c) comparison of longitudinal strains of the solitary wave with different amplitudes $H$ for several values of the dimensionless biasing electric displacement $\eta_5$ at $t_0 = 0$ and $t_0 = 5$. (Reproduced with permission from Wang et al. [107], copyright 2015 Elsevier)

In terms of strongly nonlinear waves, Wright [109,110] obtained various exact solutions of large-amplitude nonlinear waves, such as periodic and solitary waves, in an incompressible elastic material. The nonlinear traveling wave solutions were also found by Dai [111] in an elastic circular rod for an incompressible Mooney-Rivlin material with strong nonlinearity. Besides, Dai and Zhao [112] investigated nonlinear traveling waves in an elastic rod described by a modified Mooney-Rivlin material model and found seven types of nonlinear waves, including the solitary waves of radial contraction and radial expansion, kink waves, antikink waves, and three types of periodic waves. Furthermore, Dai and Li [113] examined strongly nonlinear axisymmetric waves in a circular hyperelastic rod characterized by a compressible Mooney-Rivlin material model and observed the appearance of seven types of nonlinear waves, namely solitary waves of radial contraction and radial expansion, solitary shock waves of radial contraction and radial expansion, periodic waves and two types of periodic shock waves. Recently, the large-amplitude nonlinear elastic waves in soft periodic structures were investigated [114-116] and the experimental and numerical results demonstrated the existence of the vector solitary waves in these soft periodic structures. Moreover, the results revealed that the structural geometry and the initial deformation can be harnessed to tune the wave characteristics and control the selective generation of the solitary waves.

## 5. Waves in soft DE phononic crystals

In recent years, DEs have been used to fabricate functional PCs for potential applications in tunable waveguides, filters, sensors, and many more [117-119]. Large deformations induced by external electric stimuli will significantly influence the electromechanical properties of DEs, which in turn provides an effective means to manipulate the wave characteristics of soft DE PCs, including the band gaps (BGs), negative refraction, topological interface/edge states, etc. Consequently, much effort has been devoted to the exploration of the tunability of acoustic, electroelastic, and electromagnetic wave characteristics in soft DE PCs under external electric, magnetic, and/or mechanical biasing fields, and the relevant research is summarized in this section for a convenient and fast reference.

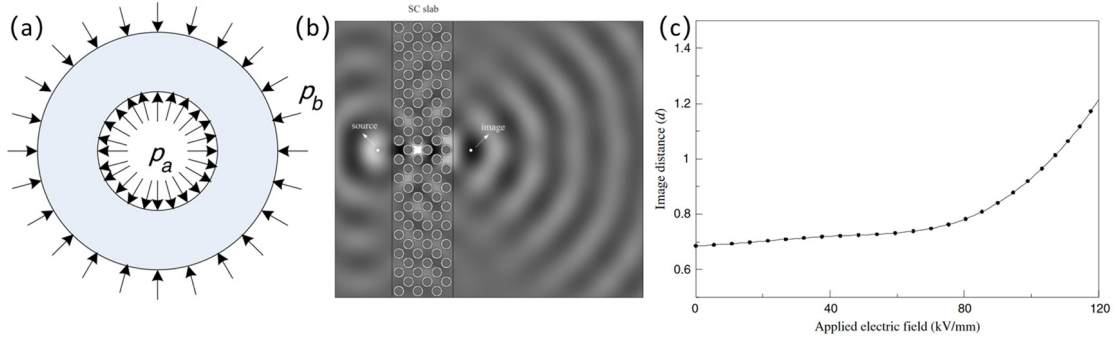

**Figure 16.** Tunable sonic crystal (SC) composed of DE cylindrical actuators with focusing function[120]: (a) Cross-section of a DE cylindrical actuator under the electrostatic pressures exerted by the electrodes when subjected to an electric biasing field; (b) the acoustic field distribution for the SC slab with focusing function; (c) the relation between the image spot position and the applied electric field. (Reproduced with permission from Yang et al. [120], copyright 2008 IOP Publishing)

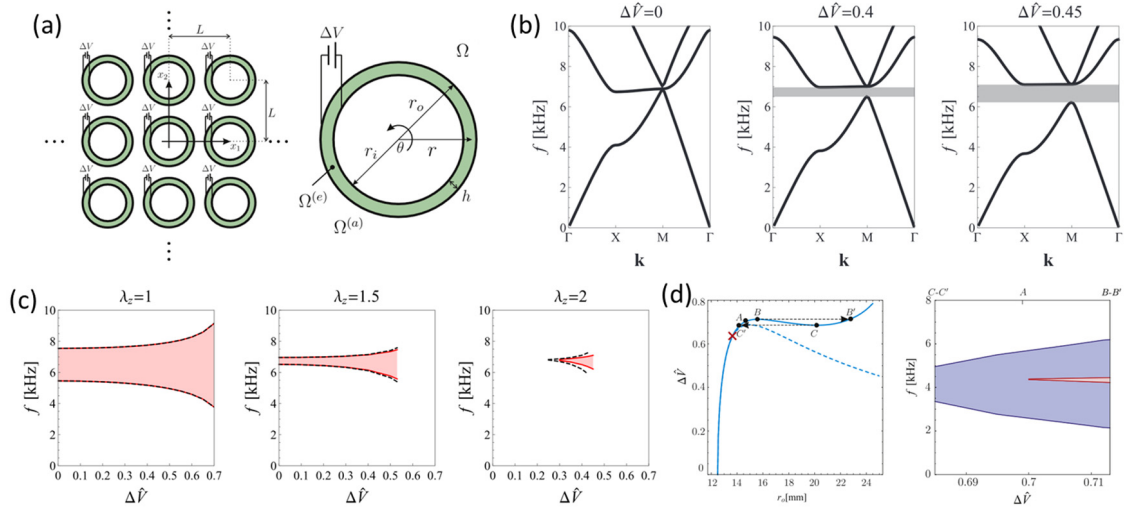

**Figure 17.** Tunable sonic BGs of a soft DE PC by adjusting the external voltage[122]: (a) Cross-section of an infinite array of actuated soft DE cylindrical tubes surrounded by air and the single DE tube in deformed state; (b) influence of the applied radial voltage on the band diagram of a square array of pre-stretch DE tubes; (c) acoustic BG as a function of the normalized applied voltage for several values of axial pre-stretch; (d) snap-through transition of the deformed outer radius with the continuous change in normalized voltage and the acoustic BG as a function of the normalized voltage for a square array of axially free DE tubes before (red region) and after (blue region) the snap-through. (Reproduced with permission from Bortot and Shmuel [122], copyright 2017 IOP Publishing)

### 5.1 Acoustic waves

The studies begin with the design of soft DE PCs with tunable acoustic properties. Yang et al. [120] designed a tunable two-phase sonic crystal (SC) consisting of the periodic DE cylindrical actuators and air by adjusting the electric biasing field. The geometric change of the DE tubes as shown in **Fig. 16**(a) induced by the electrostatic pressures exerted by the electrodes when subjected to the external electric field results in the transition of the refractive direction of the acoustic waves from positive to negative, which makes it practical to design tunable acoustic superlens with tunable focus points by adjusting the electric field (see **Fig. 16**(b) and (c)). In quick succession, Yang and Chen [121] proposed a 2D soft PC made of a square array of hollow DE cylinders surrounded by air and investigated the tunability of the acoustic wave band structures under electric loads. It was found that the external voltage could tune the width and location of the acoustic BGs. Moreover, a local BG was observed within the pass band, which enables the 2D soft DE PC to serve as an acoustic filter or switch. Recently, Bortot and Shmuel [122] designed a 2D soft PC consisting of DE cylindrical tubes immersed in an air background with tunable acoustic BGs across audible frequency range as shown in **Fig. 17**(a). The numerical results revealed that increasing the applied voltage opens and widens the acoustic wave BGs of the soft DE SC (see **Fig. 17**(b)), and the axial pre-stretch may significantly affect the width of the BGs as well (see **Fig. 17**(c)). Besides, the interesting snap-through instabilities (the outer radius of the soft DE tube changes abruptly at some particular value of the voltage) resulting from geometric and material nonlinearities were revealed, which may be explored to achieve sharp transitions in the acoustic wave BGs (see **Fig. 17**(d)). Based on the DE resonator unit in [44,45], Yu et al. [123] designed a novel acoustic metamaterial composed of a DE resonator array that was attached to the side wall of a rigid duct as shown in **Fig. 18**(a). **Fig. 18**(b) and (c) presented the fabrication of the acoustic DE metamaterial with four resonators connected in series as an array and the experimental test facilities, respectively. Because of the tunable attenuation bands of each DE resonator unit by adjusting the applied voltage, the acoustic DE metamaterial presents broadband sound attenuation performance with a voltage-controlled frequency range (see **Fig. 18**(d) and (e)), which has promising potential as tunable sound control devices.

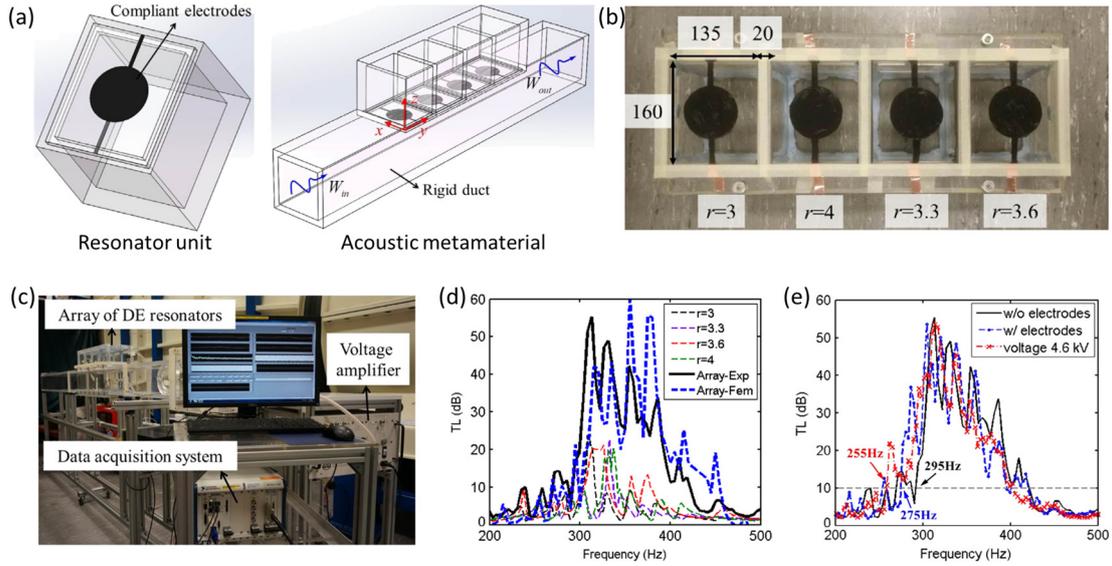

**Figure 18**. Tunable acoustic metamaterial with a DE resonator array for sound attenuation application[123]: (a) Schematics of the DE resonator unit and acoustic metamaterial; (b) fabrication of an acoustic metamaterial with four DE resonator units; (c) experimental testing facilities; (d) transmission loss spectra of the metamaterial with four DE resonator units; (e) influence of the electrodes and electric voltage on transmission loss spectra of the acoustic DE metamaterial. (Reproduced with permission from Yu et al. [123], copyright 2016 Elsevier)

### 5.2 Electroelastic waves

More attention has already been paid to the tunable electroelastic waves propagating in soft DE PC structures. To the best of the authors' knowledge, the relevant research may date back to the pioneering work by Gei et al. [124]. They theoretically demonstrated a novel way to control electroelastic wave BGs in a pre-stretched waveguide, which was made of rectangular incompressible NH DEs with the electrode pairs periodically overlaid, by applying an external voltage. The simulation results showed that the position of the BGs can be accurately manipulated by adjusting the applied voltage such that it is possible to cover the wide frequency spectrum with the steerable BGs. Different from the periodically controlled manner in Ref. [124], Shumel and Pernas-Salomón [125] proposed to manipulate the flexural waves in a two-component rectangular DE composite waveguide via electrostatically-controlled aperiodicity. Besides loading a fixed axial force, electric voltages were applied over selected DE segments to realize the aperiodically active control. The numerical simulations revealed that artificially aperiodic arrangements could actively and effect-ively control the elastic wave band structures in the DE waveguide, including the opening of new BGs and the tunability of position and width of the existing BGs. In addition, based on the transfer matrix method and the Bloch-Floquet theorem, Shmuel and deBotton [126] investigated the feasibility to control the BGs in an infinite periodic laminate composed of two types of incompressible NH DEs by adjusting the electric biasing field. The influences of material properties, volume fraction, and electric biasing field on the BGs were discussed in detail. It was found that the biasing electric displacement leads to changes in the width and location of the BGs. Moreover, the effect of the biasing electric displacement on the BGs becomes more prominent for lower contrasts between the material constants of the two phases, which enables engineers to actively tune the BGs. Galich and Rudykh [127] also investigated the shear-wave BGs (SBGs) in soft DE laminates subjected to an electric biasing field for the incompressible NH and Gent material models. They revealed that the SBGs are independent of the elastic deformation or the electric biasing field for the periodic laminate with NH phases because the change in the geometry completely cancels out the variation of phase velocity induced by the deformation, which is in contradiction with the conclusion drawn in Refs. [126,128,129]. However, for the periodic laminate with Gent phases, the electric biasing field widens the SBGs and leads to the shift of the SBGs towards higher frequencies, once the strain-stiffening effect is triggered. Zhu et al. [130] further studied the anti-plane SH waves propagating in a periodic laminate composed of the Gent DEs with oblique incidence, and obtained the band structures and energy transmission spectra under different external electric loads. The calculated results showed that the distributions of the BGs can be actively controlled by the electric biasing field and the incident angle. Moreover, only if the number of the unit cell is large enough, the SH wave transmission ratio of finite periodic laminates can be used to predict the BGs of the corresponding infinite periodic structure. More generally, Chen et al. [131] considered the shear and longitudinal waves propagating in a compressible DE laminate as shown in **Fig. 19**(a) and comparatively investigated the influences of different energy function models including the electrostriction on the BGs of a soft DE laminate subjected to the electrostatic voltage and prestress. The snap-through instability of the DE laminate for the Gent model resulting from geometric and material nonlinearities was revealed (see **Fig. 19**(b)), which can be used to achieve a sharp transition in the width and location of the BGs for both the shear and longitudinal waves (see **Fig. 19**(c)). However, the snap-through transition does not appear in the band structures for the NH model or Gent model with a strong strain-stiffening effect (see $J_m = 5$ in **Fig. 19**(b)) such that the BGs could be continuously tuned by adjusting the electric biasing field. The numerical results also indicated that although both the electrostrictive effect and prestress contribute to stabilizing the periodic DE laminate, they influence the BGs in different manners: the former weakens the BG jump induced by the snap-through transition (see **Fig. 19**(d)), while the latter broadens the tunable frequency range for both shear and longitudinal waves (see **Fig. 19**(e)). In addition, Wu et al. [132] designed a 1D PC composed of aluminum and polymethyl methacrylate layers with a DE defect layer subjected to an electric biasing field and observed the defect bands in the BGs.

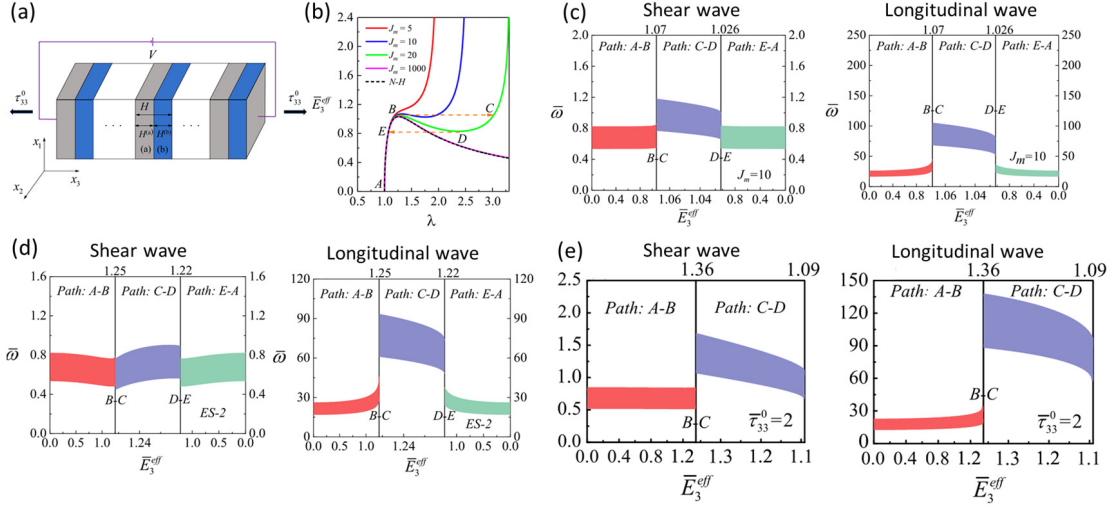

**Figure 19.** Tunable elastic waves in compressible DE laminates subjected to electric voltage and prestress[131]: (a) An infinite two-phase periodic DE laminate subjected to voltage and prestress along the thickness direction; (b) lateral stretch as a function of the normalized effective referential electric field in the DE laminate for the Gent model with different Gent constants and the NH model; (c) the first BG of incremental shear and longitudinal waves versus the normalized electric field in the laminate of the Gent model with $J_m$=10; (d) the first BG of incremental shear and longitudinal waves versus the normalized electric field in the laminate of the Gent model ($J_m$=10) with electrostriction effect; (e) the first BG of incremental shear and longitudinal waves versus the normalized electric field in the laminate of the Gent model ($J_m$=10) with prestress. (Reproduced with permission from Chen et al. [131], copyright 2020 Elsevier, Open Access)

Besides, the frequencies of the defect bands could be tuned by adjusting the applied voltage, which may be explored to develop electrostatically-controlled narrow passband filters.

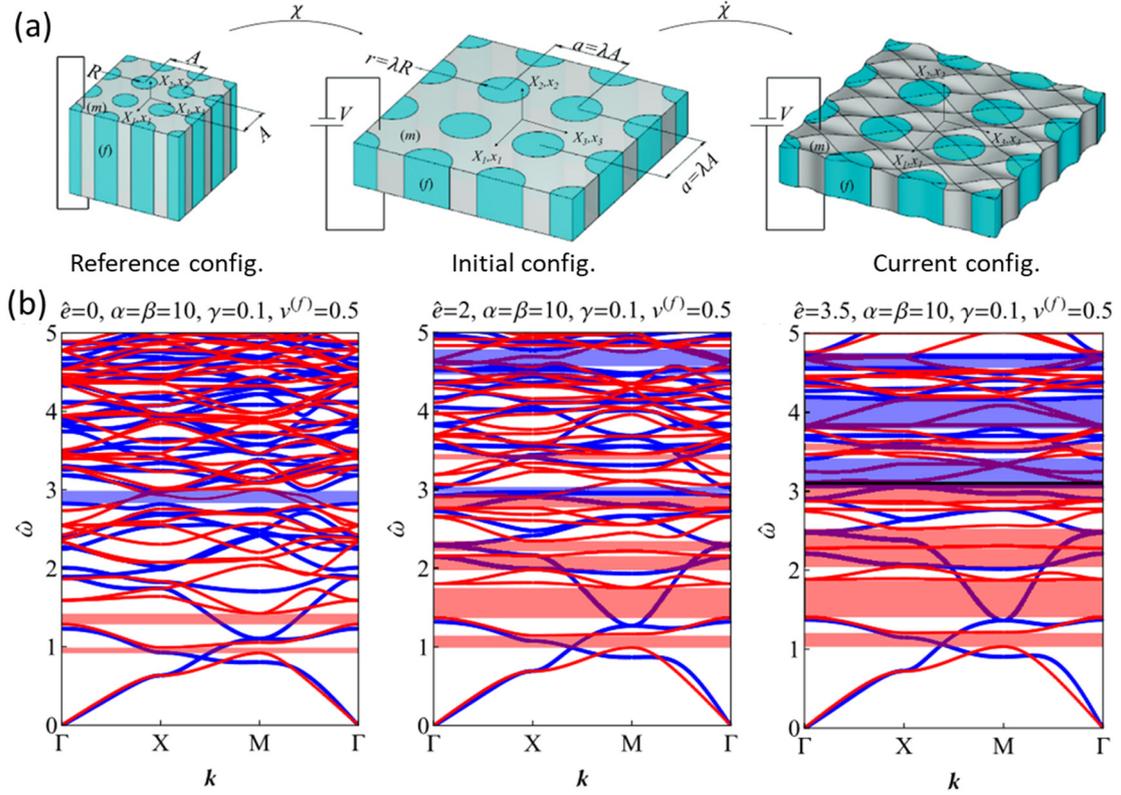

**Figure 20.** Voltage-controlled complete BGs in a 2D fiber composite made of DEs[134]: (a) Fiber composites in the undeformed reference configuration, initial configuration, and current configuration; (b) band diagrams of the 2D DE fiber composite subjected to different electric biasing fields. The blue and red curves correspond to the normalized frequencies of the in-plane and anti-plane wave motions, respectively. The blue, red, and black regions correspond to the in-plane, anti-plane, and complete BGs, respectively. (Reproduced with permission from Getz et al. [134], copyright 2016 Elsevier)

Apart from the laminate-based PCs, there also exist novel PC structures made of cylindrical DEs. For example, Shmuel [133] designed a fiber-reinforced DE composite PC with square lattice and studied band structures of the anti-plane waves propagating in the soft PC subjected to an electric biasing field along the DE fibers, whose constitutive behavior is characterized by the incompressible Gent model. It was found that the BG properties of the anti-plane waves in the fiber-reinforced composite are significantly influenced by the material contrast, matrix volume fraction, and the electric biasing field. Most importantly, enhancing the electric field intensity increases the width and frequency of the BGs. Recently, Getz et al. [134] considered the same fiber-reinforced composite PC (see **Fig. 20**(a)) as the one designed by Shumel [133], and extended the results in Ref. [133] by considering the deformation-dependent permittivity of DE materials. In addition to the conclusions drawn by Shumel, the main finding was the existence of complete wave BGs, which are independent of the propagation direction and the plane of the motion, for some particular compositions and electric biasing fields (see **Fig. 20**(b)). Later, Getz and Shmuel [135] also proposed an incompressible Gent DE fiber composite plate with voltage-controlled BGs by adjusting the applied voltage. They found that the electrostatically-tunable wave BGs strongly depends on the thickness of the composite plate. Besides, for certain combinations of plate thickness, volume fraction of the constitutions, and material properties, the complete wave BGs in the composite plate can be achieved with a prominently lower electric stimulus than the one required in the bulk counterpart [134]. Different from the above-mentioned studies based on theoretical methods, Jandron and Henann [136,137] developed a finite-element-based numerical method for designing electrostatically-tunable soft DE PCs. By applying the electric voltage parallel to or perpendicular to the fibers, they presented the tunable band structures of the 2D PCs composed of circular-cross-section fibers embedded in a matrix with rectangular lattice or hexagonal lattice. The numerical results indicated that the number, width, and location of the BGs in the soft DE PC can be significantly affected by the material parameter contrast and electric loading direction. The electrical tunability of the BGs can be further enhanced by the electromechanical pull-in instability and the large-stretch chain-locking behavior. In addition, a 1D DE phononic cylinder with longitudinal wave modes was proposed by Wu et al. [138], where the axial mechanical load (including two loading paths, i.e., fixed axial pre-stretch (path A) and fixed axial force (path B)) and the axial voltage applied to the periodic electrodes attached to the ends of every DE cylinder segment were used to control the band structures of the DE PC cylinder for both NH and Gent models. The numerical results revealed that both large deformation and electromechanical coupling enable one to actively manipulate the BGs in soft DE PCs. More specifically, increasing the applied voltage can open and significantly widen the BGs, while the axial pre-stretch or axial force changes the position and width of the BGs. Besides, the axially free DE PC cylinder described by the Gent model presents the snap-through instability, which helps to realize sharp transitions of the BGs.

Recently, some research groups reported novel soft DE PCs with artificially designed unit-cell structures based on the topology optimization so that the resulting PCs achieve desired properties and functionalities. For example, by adopting the genetic algorithm (GA), which is recognized as a kind of non-gradient topology optimization (NGTO) method, Bortot et al. [139] maximized the width of the anti-plane wave BGs in tunable DE PCs subjected to a prescribed electric field, in which two different representations of the topology for the unit cells, i.e., pixel grid and B-spline, were used as the design variables. Optimization results indicated that both the pixel grid-based and B-spline-based descriptions demonstrated an improvement in the BG width by adopting a unit-cell structure with an optimal circular fiber, while the required number of design variables for the B-spline-based description is much smaller than the one for the pixel grid-based description. Furthermore, to overcome the limitation of design variables in classical GAs, Sharma et al. [140] developed an efficient gradient-based topology optimization (GTO) method, namely the Method of Moving Asymptotes (MMA), for maximizing the width of the anti-plane shear wave BGs in tunable soft DE PCs based on the finite element formulation. They proved that the MMA identifies the microstructural unit cells with wider BGs than the GA, while the computational cost to reach these optimal unit cells is reduced by orders of magnitude. Although the GTO methods (e.g., the MMA-based framework) exhibit higher computational efficiency than the NGTO methods, they more depend on the selected initial designs with existing BGs. Very recently, Hu et al. [141] proposed a novel gradient-based optimizer, i.e., the Trust Region based Moving Asymptotes (TRMA), which showed stronger robustness than the frequently used MMA in solving some benchmark topology optimization problems (like stress-constrained problems). Therefore, the newly developed TRMA may provide a potential strategy to solve the initial-value dependency problem involving in the traditional GTO method for the soft DE PCs.

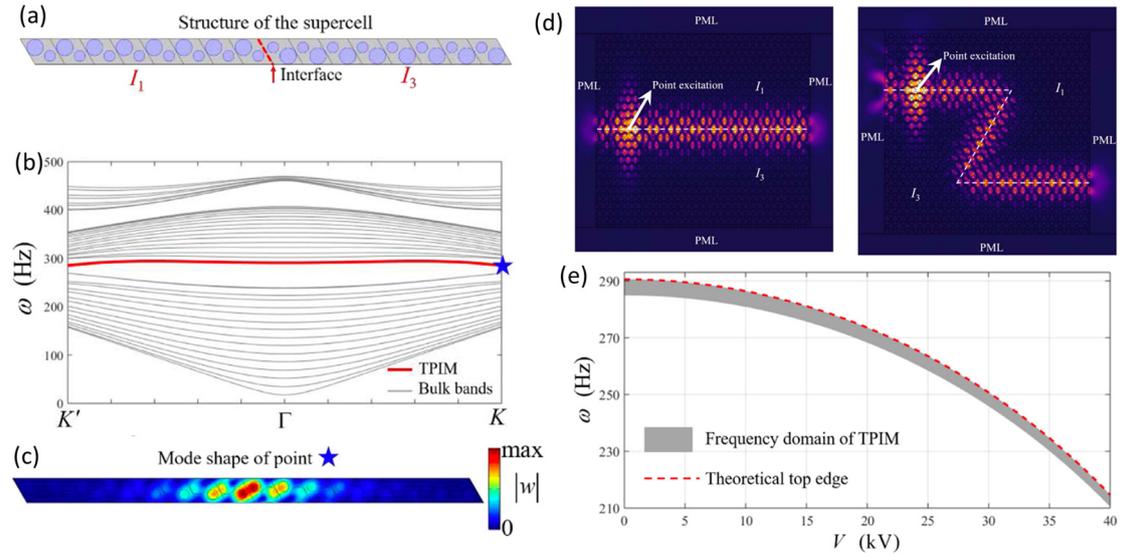

**Figure 21.** Voltage-controlled topological interface modes in DE acoustic metamaterials[146]: (a) Supercell of DE membrane-type acoustic metamaterials (MAMs) composed of two types of topological phases with the interface denoted as the red dashed line; (b) band structure of the composite MAM; (c) the mode shape of the topologically protected interface mode (TPIM); (d) two types of waveguides with the straight and zig-zag paths; (e) influence of the electric voltage on the frequency-domain of TPIM. (Reproduced with permission from Zhou et al. [146], copyright 2019 Elsevier)

At the end of this subsection, we pay attention to a few interesting and practical topics about soft PCs. First, in contrast to the aforesaid tunable soft PCs via external electric loads, one wishes to keep the BG structures unchanged, i.e., the wave BGs are independent of the imposed deformation, which enables engineers to design special soft PCs with robust performance. The idea has been realized by Zhang and Parnell [142], and the underlying necessary condition to achieve this unique property, namely the Lagrangian elasticity tensors (LETs) of the materials should be independent of deformations, was revealed by them. It was demonstrated that only incompressible NH materials satisfy this condition exactly under all deformations. Besides, semilinear hyperelastic materials could exhibit almost-constant LETs for certain deformations and material constituents. Second, different from the tunable Bragg-scattering DE PCs, two kinds of 2D locally resonant PCs with electrostatically-tunable BGs were proposed by Zhou and Chen [143]. It was found that enhancing the electric field intensity widens the BGs and leads to the shift of BGs towards higher frequencies. Moreover, the initial stress significantly affects the opening, closing, and shift of wave BGs, which enables these locally resonant PCs to be exploited as an elastic wave transition switch. Third, the topological insulator concept has been employed to manipulate the wave propagation behaviors in PCs or metamaterials. In these topological acoustic structures, the wave in the topological states can propagate along the edge or interface only [144,145] which can be used to tailor the wave propagation path with immunity from back-scattering. Very recently, Zhou et al. [146] designed a membrane-type acoustic metamaterial (MAM) composite constructed by combining state $I_1$ MAM and state $I_3$ MAM as shown in **Fig. 21**(a), and studied the tunability of the topologically protected interface mode (TPIM) when subjected to external electric loads. They found that in the complete BG of the composite MAM, there exists the unique TPIM marked by the red line as shown in **Fig. 21**(b) with its wave energy trapped around the interface of two topological phases (see **Fig. 21**(c)). Further, they designed two types of waveguides that exhibit the straight path and zig-zag path, respectively, and the most input energy is translated into the TPIM (see **Fig. 21**(d)). Moreover, it was found that the applied voltage sensitively tunes the frequency domain of TPIM over a wide range (see **Fig. 21**(e)) without influencing its localization behavior, which can be explored to achieve frequency-tunable waveguides with directional propagation paths. Last, magnetorheological PCs [147-149], as an interesting and significant topic, are composed of soft magnetoactive materials which are achieved by mixing the elastomer materials with microsize magnetic particles. When applying a magnetic biasing field, the soft magnetoactive materials can deform and change the effective material properties. Some research groups have reported the designs of soft magnetorheological PCs [150-152], and the magnetic-field controlled wave BGs were examined when subjected to the magnetic biasing fields.

### 5.3 Electromagnetic waves

The DE photonic crystals are briefly introduced here as a prolongation. Photonic crystals, as artificial structures similar to PCs, are mainly used to modulate the propagation characteristics of light waves (or electromagnetic waves), and some unique optical properties like negative refraction and subwavelength imaging were realized [153-155]. Recently, novel photonic crystals containing DE components were designed and their electrostatically-controlled optical functions were examined. For example,

Wang and Chen [156] proposed a tunable directional coupler constructed by a 2D photonic crystal made of DE cylindrical tubes in air background, and the tunable output power of light waves was exhibited. The 2D graded photonic crystal composed of a DE cylindrical tube array was proposed by Wang et al. [157] to integrate two functions, i.e., light focusing and beam-steering, into one structure. They found that its focal length and beam deflection angle could be actively tuned by controlling the electric voltage applied to the DE tube array.

## 6. Conclusions and outlooks

### 6.1 Conclusions

This paper provides a review of the recent advances in the vibrations and waves in soft DE structures of the plate, cylindrical, and spherical configurations as well as DE phononic crystals and metamaterials. Based on the representative nonlinear continuum theory formulated by Dorfmann and Ogden, which is substantially equivalent to other versions of nonlinear continuum mechanics frameworks, small-amplitude incremental vibrations and waves superimposed on finitely deformed DE structures of three types of configurations under mechanical and/or electric loads have been investigated in detail. The nonlinear vibrations and waves in soft DE structures subjected to constant and time-dependent excitations have also been summarized for the completeness of the article. A general conclusion can be drawn that the large deformation, electromechanical coupling, and external electric stimuli have significant influences on the vibration and wave characteristics of soft DE structures, which can be harnessed to realize tunable DE multi-functional structures and devices. In comparison with the past researches, remarkable advances have been achieved in recent years and are summarized below:

(1) Small-amplitude vibrations and waves in finitely deformed soft DE structures of the plate (rectangular and circular), cylindrical (solid and hollow), and spherical (hollow) configurations have been systematically investigated when subjected to mechanical and/or electric loads. Plentiful deformation patterns including the shear, flexural, longitudinal, circumferential, non-axisymmetric, torsional and spherical modes are exhibited. The prominent influences of large deformation, mechanical force, and electric biasing field on vibration and wave characteristics have been revealed in detail.

(2) Tunable DE PCs and acoustic metamaterials composed of various substructures, such as periodically heterogeneous laminates, membrane-based structures with periodically electric conditions, composites consisting of cylindrical DE tubes and matrix (or air), cylindrical DEs with periodically electric conditions, and novel unit cells designed by topology optimization methods, have been designed by adjusting external mechanical and electric loads. Some unique PC materials and structures have been proposed to realize special functions, for example, wide tunable phononic BGs, the band structure robust to the deformation, topologically protected unidirectional wave propagation, etc.

(3) Nonlinear vibrations in finitely deformed soft DE structures of plate, cylindrical, and spherical configurations subjected to time-dependent mechanical or electric excitations have been investigated. The influences of material viscoelasticity, geometric and material nonlinearities, and external excitations on the nonlinear frequency and oscillation amplitude have been discussed in detail. Moreover, the weakly nonlinear elastic waves propagating in finitely deformed DE cylinders have also been studied. The existence of nonlinear solitary waves in DE cylinders has been theoretically demonstrated.

## 6.2 Outlooks

Despite substantial achievements that have been made, the significant issues listed in the following remain to be addressed in future studies.

(1) It has been found that material viscoelasticity significantly affects the material properties of DEs [158], the nonlinear responses of soft DE structures [78-83], and band structures of soft elastic PCs [159]. However, the influence of material viscoelasticity is usually neglected during the study on small-amplitude vibrations and waves in soft DEs as well as DE PCs and metamaterials, which should be further explored for comprehensive prediction and understanding of the dynamic behaviors of soft DE structures and devices.

(2) Most of the works on tunable DE PCs and metamaterials are based on theoretical calculations or numerical simulations. The reports on experimental fabrications of soft DE PCs and metamaterials, especially with active or smart control capability, are still limited, which restricts the real applications of the DE PC-based devices. The potential solution to the high-precision shaping and patterning of PCs and metamaterials is the novel 3D printing techniques [160-162]. Therefore, developing fast and high-precision fabrication techniques for DE PCs and metamaterials is a key and challenging issue in the future.

(3) A few studies have reported the tunable vibration and wave characteristics of soft DEs as well as the tunable band structures of soft DE PCs realized by the snap-through instabilities when applying the fixed tensile forces. However, when applying compressive forces, the DEs and DE PCs will be susceptible to buckling and post-buckling instabilities [163-166]. Therefore, the influence of instabilities induced by compressive forces on the small-amplitude vibrations and waves in soft DEs as well as the band structures in DE PCs needs to be further studied. Moreover, tuning the vibration and wave characteristics in soft DEs and DE PCs through buckling instabilities could be another interesting topic for future research.

(4) In most reported works on nonlinear vibrations, the applied mechanical and electric loads are described by deterministic and specific functions such as the sinusoidal function or the constant excitation pulsing the sinusoidal excitation. During practical operation, DE structures will be susceptible to random disturbance, and thus the random vibration responses of DE structures should be further explored. Moreover, the influence of time-dependent mechanical and electric excitations perturbed by a random disturbance on the dynamic responses of DE structures is also a meaningful research topic.

(5) Fluid-solid interaction is a significant and challenging issue due to the potential applications of DE structures and devices in ocean exploration [167,168], biological medicine [169,170], and so on. The mechanical models of fluid-solid interfaces [170,171] remain to be established to describe the interactions between the DE structures and different fluids (Newtonian fluids, non-Newtonian fluids, etc). The influences of mechanical and electric biasing fields along with nonlinearity on the dynamic characteristics of DE structures surrounded by fluids should also be investigated. Active and smart control of vibrations and waves in fluid environments is also desired.

(6) It has been proven that the inhomogeneous material properties, electric biasing fields, and pre-stretch ratios can be harnessed to manipulate the vibration characteristics of pre-stretched DE membrane-based structures. The influences of inhomogeneous (or even graded) material properties as well as inhomogeneous mechanical and electric biasing fields on vibration responses and wave propagations of soft DE cylindrical and spherical structures, which have not been explored yet, will be more complicated, but it is also an interesting topic in future studies.

(7) Different from the reported DE PCs, the porous periodic DE structures, as a special design, could be another potential scheme to tune the BG properties. Porous periodic structures are easy to be manufactured [172-174] and the holes can also be filled with other materials [159]. Besides mechanical or electric biasing fields, the shape and periodic distribution of the holes as well as the inclusions can be exploited to manipulate the position and width of the BGs of porous DE PCs.

(8) The nondestructive testing method based on nonlinear waves possesses superior advantages such as fast detection speed, high sensitivity, high precision, etc [175-177]. For DE materials, it has been revealed that the electric biasing field can tune the wave shape and wave velocity of the nonlinear solitary waves in real time [107], which can be explored to detect or manifest the defects or fatigue cracks by exciting the nonlinear waves with different wave velocities. However, existing studies on nonlinear waves in soft DE materials/structures remain limited. Therefore, the propagation characteristics of various types of nonlinear waves in soft DE structures (or even DE PCs) subjected to mechanical and electric biasing fields need to be further investigated. Moreover, the effects of competition between nonlinearity, dispersion, viscoelasticity, and biasing fields on nonlinear wave propagation are worth in-depth exploration.

(9) The design of the periodic unit cell of PCs for desired performance is a typical topology optimization problem [178]. Various optimization methods including the GTO [179-182] and NGTO [183,184] have been proposed to seek the optimal material distributions within the unit-cell structure of 2D and 3D elastic PCs. However, due to the electromechanical coupling and large deformation, topology optimization for designing periodic unit cell of 2D and 3D soft DE PCs is more complex and the related researches are still very limited [139,140]. Therefore, developing effective topology optimization methods to handle problems with electromechanical coupling and geometric and material nonlinearities is also a meaningful and significant topic for the real applications of active DE PC devices.

## Acknowledgments

The work was supported by the National Natural Science Foundation of China (Nos. 11872329, 12072315, 12192211), the Natural Science Foundation of Zhejiang Province (No. LD21A020001), and the National Postdoctoral Program for Innovative Talents (BX2021261). Bin Wu acknowledges the support of the European Unions Horizon 2020 Research and Innovation Programme under the Marie Skodowska-Curie Actions (No. 896229). Partial support from the Shenzhen Scientific and Technological Fund for R&D, PR China (grant number 2021Szvup152) is also acknowledged.